\documentclass[aps,prd,reprint,groupaddress,noshowpacs,noshowkeys]{revtex4-1}

\usepackage{amsmath,amssymb,mathrsfs}

\usepackage{graphicx}% Include figure files
\usepackage{dcolumn}% Align table columns on decimal point
\usepackage{bm}% bold math
\usepackage{xcolor}
\usepackage[T1]{fontenc}
\usepackage[pdftex,
  hidelinks,
  colorlinks=true,
  linkcolor=black,
  anchorcolor=black,
  citecolor=black,
  urlcolor=black,
  pdfauthor= {Michael Glinsky},
  pdfsubject = {Mallat Scattering Transformation},
  pdfdisplaydoctitle,
  bookmarks=true,
  bookmarksopen=true]{hyperref}

\begin{document}
\title{Quantification of MagLIF Morphology using the Mallat Scattering Transformation}

\author{Michael E. Glinsky}
\affiliation{BNZ Energy Inc., Santa Fe, NM, USA}

\author{Thomas W. Moore, William E. Lewis, Matthew R. Weis, Christopher A. Jennings, David A. Ampleford, Eric C. Harding, Patrick F. Knapp, Matthew. R. Gomez and Sophia E. Lussiez}
\affiliation{Sandia National Laboratories, Albuquerque, NM, USA}

%\date{\today, DRAFT, version 1}

\begin{abstract}
The morphology of the stagnated plasma resulting from Magnetized Liner Inertial Fusion (MagLIF) is measured by imaging the self-emission x-rays coming from the multi-keV plasma, and the evolution of the imploding liner is measured by radiographs. Equivalent diagnostic response can be derived from integrated rad-MHD simulations from programs such as Hydra and Gorgon. There have been only limited quantitative ways to compare the image morphology, that is the texture, of simulations and experiments. We have developed a metric of image morphology based on the Mallat Scattering Transformation (MST), a transformation that has proved to be effective at distinguishing textures, sounds, and written characters. This metric has demonstrated excellent performance in classifying ensembles of synthetic stagnation images. We use this metric to quantitatively compare simulations to experimental images, cross experimental images, and to estimate the parameters of the images with uncertainty via a linear regression of the synthetic images to the parameters used to generate them.  This coordinate space has proved very adept at doing a sophisticated relative background subtraction in the MST space.  This was needed to compare the experimental self emission images to the rad-MHD simulation images.
    
We have also developed theory that connects the transformation to the causal dynamics of physical systems.  This has been done from the classical kinetic perspective and from the field theory perspective, where the MST is the generalized Green's function, or S-matrix of the field theory in the scale basis.  From both perspectives the first order MST is the current state of the system, and the second order MST are the transition rates from one state to another.
    
An efficient, GPU accelerated, Python implementation of the MST was developed.  Future applications are discussed.
\end{abstract}

\maketitle

\onecolumngrid

\section{Executive Summary}\label{sec:summary}
The morphology of the stagnated plasma resulting from Magnetized Liner Inertial Fusion (MagLIF) is measured by imaging the self-emission x-rays coming from the multi-keV plasma, and the evolution of the imploding liner is measured by radiographs. Equivalent diagnostic response can be derived from integrated rad-MHD simulations from programs such as Hydra and Gorgon. There have been only limited quantitative ways to compare the image morphology, that is the texture, of simulations and experiments. We have developed a metric of image morphology based on the Mallat Scattering Transformation (MST), a transformation that has proved to be effective at distinguishing textures, sounds, and written characters. This metric has demonstrated excellent performance in classifying ensembles of synthetic stagnation images. We used this metric to quantitatively compare simulations to experimental images, cross experimental images, and to estimate the parameters of the images with uncertainty via a linear regression of the synthetic images to the parameters used to generate them.  This coordinate space has proved very adept at doing a sophisticated relative background subtraction in the MST space.  This was needed to compare the experimental self emission images to the rad-MHD simulation images.
    
We have also developed theory that connects the transformation to the causal dynamics of physical systems.  This has been done from the classical kinetic perspective, where the MST are expected values of the generalized Wigner-Weyl transformations of the density operator.  And, has been done from the field theory perspective, where the MST is the generalized Green's function, or S-matrix of the field theory in the scale basis.  From the classical perspective, the first order MST is the one particle distribution function, averaged over the fast dynamical time scale, and the second order MST is simply related to the fully nonlinear transition rate from one scale to another.  The first gives the current state, and the second gives the nonlinear evolution of the system.  From the field theory perspective, the first order MST is the classical action averaged over fluctuations as a function of the inverse scale, and the second order MST is the scattering cross section from an initial to a final scale.  The first again gives the current state of the system, and the second gives the inverse mass of the field boson that mediates the field interaction and scatters the field, thereby evolving the field.  What is required of the system is that it is causal.  Equivalently, a Lagrangian for the system can be written down, and therefore the system will be evolved according to an action principle.  This leads to a generalized advection by a Lie derivative, in the classical case, and by dynamical paths weighted by the exponential of their actions, in the field theory case.
    
Therefore, it is no surprise that the MST is a good metric for nonlinear systems.  It encodes both the initial state and the dynamics (transition rates) between states.  This explains why the first order MST, or the Fourier transformation, are not sufficient to uniquely identify the systems.  They only encode the the initial state.  There are other technical details that cause problems with the Fourier Transform, unless there are rotational symmetries in the physical system.  It is also the reason a system that encodes finite information is fully identified by the first and second order MST.  This comes about because of a statistical realizability theorem  -- either the distribution stops at second order or it must continue to all orders.  Since the information is finite, the distribution must stop at second order.  One can reason to this since knowing the S-matrix, that is the MST, is equivalent to knowing the Lagrangian of the field theory.
    
What is even more important about the MST is the connection to the dynamical evolution of the physics.  If one includes the evolution coordinate, that is time, in the transformation, the second order MST directly, and with no further transformation, gives the transition kernel of the dynamics.  This is independent of the current state, that is the first order MST.  Given an ensemble of example states that sufficiently sample the transition kernel, one has fully characterized the physical system and should be able to evolve any state forward in time, as given by the initial first order MST.  That is the MST is the perfect coordinate system in which to learn, identify, and propagate the dynamics.
    
The MST has been implemented in an efficient (GPU accelerated), yet flexible, Python framework based on Keras/Tensorflow.  This package includes 1D, 2D and 3D transformations along with visualization.  It supports, through adjoints, the inversion of the transformation.  The software is open source and distributed through PyPI as the BluSky project.
    
Future work that builds upon this fundamental work could include use of the MST as:  (1) a metric in the objective function of a Bayesian data assimilation,  (2) a coordinate system in which to numerically integrate physical dynamics, (3) a coordinate system in which to machine learn how to numerically integrate physical dynamics, (4) a metric to identify phase transitions, that is bifurcations, in the physical dynamics, and (5) a coordinate system in which to build surrogate models.

\section{Introduction}\label{sec:intro}
Magnetized Liner Inertial Fusion (MagLIF) is a magneto-inertial fusion concept currently being explored at Sandia's Z Pulsed Power Facility \citep{Slutz2010,Awe2013,Gomez2014}. MagLIF establishes thermonuclear fusion conditions by driving mega-amps of current through a low-Z conducting liner. The subsequent implosion of the liner containing a preheated and premagnetized fuel of deuterium or deuterium-tritium compresses and heats the system, creating a plasma with fusion relevant conditions.

Developing a detailed understanding of how experimental parameters such as premagnetization, preheat, and liner design mitigate losses and control the ignition, as well as evolution of the plasma, is a crucial and ongoing step towards realizing the full potential of MagLIF. To this end, time resolved radiography of the imploding liner, as well as self emission x-rays from the plasma at stagnation (where thermal pressure of the plasma stalls the liner implosion) have been used to study the evolution of the plasma and its structure at peak fusion conditions. For example, \citet{Awe2013} observed an unexpected feature in radiographs of a magnetized imploding liner -- a double helical structure not observed in non-magnetized liners. Additionally, bifurcated double helical strands have been observed in the stagnating plasma columns captured by self emission x-ray image diagnostics.

The underlying physics linking the double helix structure of the imploding liner to the bifurcated double-helices in the stagnated plasma is as of yet unknown. One working hypothesis is that a helical magnetic Rayleigh-Taylor instability (MRT) \citep{Seyler2018} seeded on the outside liner surface may grow large enough to feed-through the liner to seed perturbations on the liner interior. It is thought that these interior perturbations may imprint the double helical structure on the plasma. It has been experimentally demonstrated that the helical structure is dependent on the aspect ratio of the liner (AR $\equiv$ initial liner outer radius$/$initial liner wall thickness) \citep{McBride2012}. Recent experiments with varying liner thicknesses appear to demonstrate that in the case of uncoated liners the stagnation column helical radius increases while helical wavelength decreases with increasing AR, which is consistent with MRT feed-through from from the outer liner surface \citep{Ampleford2019}.   There is another working hypothesis that this double helical structure might be an emergent structure of the nonlinear evolution of the MRT that is controlled by conserved magnetic and cross helicities that are injected into the liner.  The large scale self organization would be the result of a Taylor relaxation \citep{taylor1986relaxation}, that is an energy minimization under the constraints of the topologically conserved helicities.  However, such inferences remain weak due to the fact that, to date, there has been no systematic way to quantitatively compare stagnation morphology experiment-to-experiment or experiment-to-simulation while accounting for the uncertainty in characterizing features such as the helical wavelength and radius.

In this work, we develop a method which enables such a comparison by applying a cutting edge machine learning (ML) algorithm in image classification known as the Mallat Scattering Transform (MST) \citep{Mallat2012,Bruna2013}. Particularly, we are able to use the MST to aid in inferring morphological features with uncertainty quantification. In Sec.~\ref{sec:theory}, we motivate the use of and supply the required theory for the MST. Section~\ref{sec:application} describes the synthetic model used to parameterize the double helix morphology.  We then introduce a preliminary classification model which demonstrates the ability of the MST to distinguish between different types of helical images. We close Sec.~\ref{sec:application} with details of the full machine learning pipeline used to quantify the morphological parameters of the double helical images with uncertainty. Section~\ref{sec:results} demonstrates the application of the method in quantitatively comparing simulation and experiment, as well as a direct extraction of the morphological parameters with uncertainty from experimental images. Particularly, we highlight the viability of the method to differentiate between plasmas produced from different experimental designs.  Section~\ref{sec:interpretation} develops the theory that relates the MST to causal dynamics of physical systems -- both classical and field theoretical.

\section{Mallat Scattering Transform}\label{sec:theory}
Recently, the use of deep learning methods, combined with availability of large labeled data sets, has enabled a revolution in the field of image classification and analysis.  Particularly, convolutional neural networks (CNNs) have gained widespread popularity for image analysis problems, such as classification \citep{LeCun1989}, segmentation \citep{Ning2005}, and even image generation \citep{Goodfellow2014}. The ubiquity of this approach is largely based on the ability of CNNs to learn convolutional filters which compute features that are approximately invariant to irrelevant symmetries present in the task (\textit{e.g.} translation or rotational symmetries) \citep{LeCun2010}. 

However CNNs require significant expertise to navigate a seemingly arbitrary design space (e.g. number of nodes and layers) and require considerable computing resources to train, even when using transfer learning. Additionally, their {\textit{black box}} nature make CNNs a less attractive framework for scientific applications to bridge the gap between causation and correlation. Alternative kernel classifiers such as the probabilistic neural network, are based on the Euclidean distance between image features (e.g. pixel information), which is easily broken by transformations, rotations and scaling. At the same time, familiar translation invariant feature representations such as the Fourier transform modulus are unstable to deformations (that is not Lipschitz continuous). The wavelet transformation on the other hand, is Lipschitz continuous to deformation, but is not translation invariant \citep{Bruna2013}. By combining local translation invariance and Lipschitz continuity to deformations in a fixed weight convolutional network, the MST addressed many of the concerns that arise in deep learning \citep{Mallat2012,Bruna2013}. 

MST consists of compositions of wavelet transformations coupled with modulus and non-linear smoothing operators which form a deep convolutional network. Unlike deep convolutional neural networks, the filters in the MST are prescribed rather than learned. In fact the deep convolutional network of the MST has been shown to outperform CNNs for image classification tasks over a broad range of training sample sizes \citep{Bruna2013}. This is most significant when the amount of training samples is considerably limited \citep{Bruna2013}, which is often the case with experimental data. Additional benefits of the MST framework over CNNs come in the form of intelligible design -- for example the depth of an MST network is bound by a signal's energy propagation through the network, whereas the depth of a CNN is seemingly arbitrary.

\begin{figure}[t!]
\center\includegraphics[width=25pc]{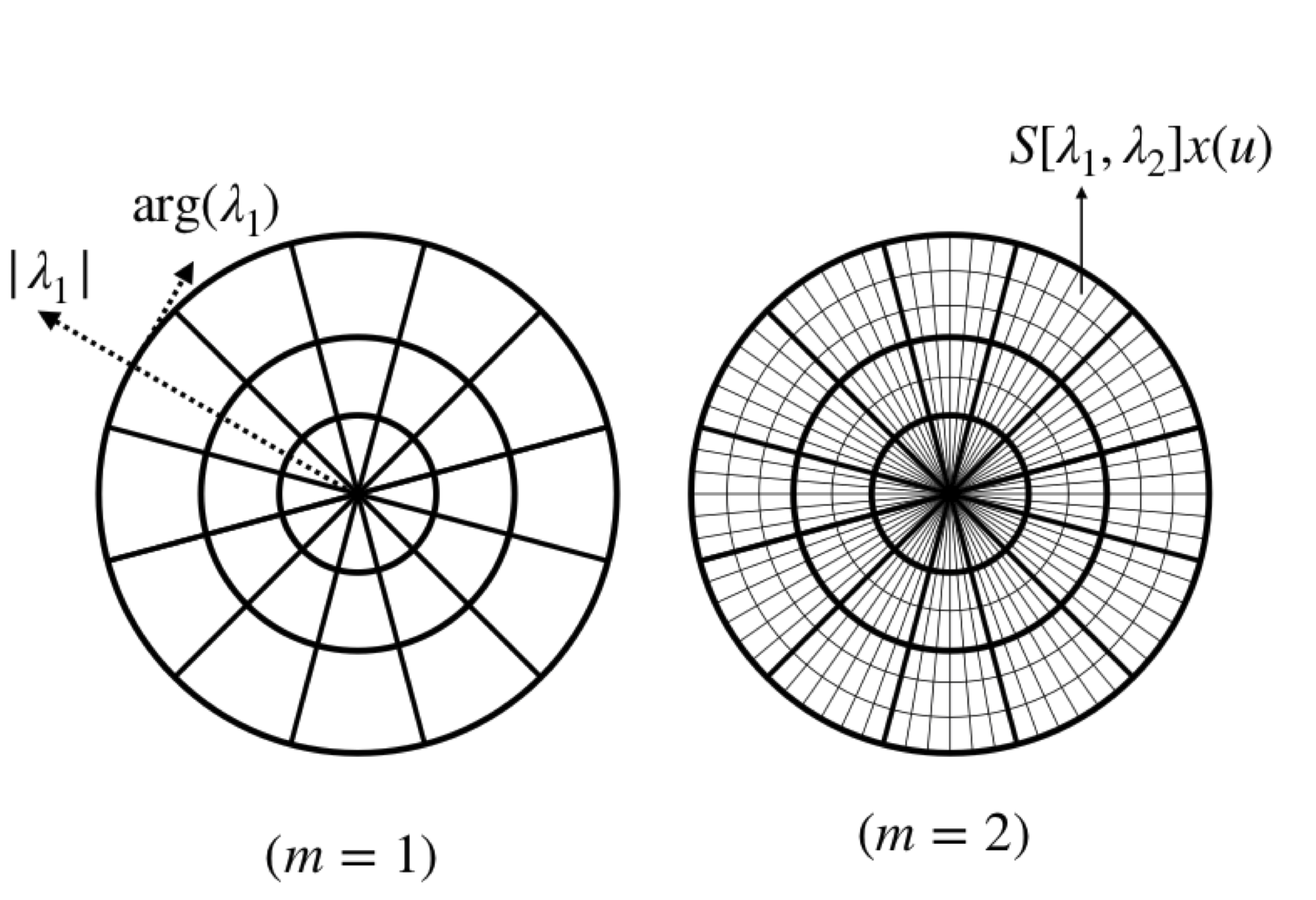}
\caption{\label{fig:logpolar} Coefficients produced by applying MST to 2D images in this work will be displayed on radial plots as shown. Bins are created according to scale (radial positioning $|\lambda_m|$) and rotation (theta position $\text{arg}(\lambda_m)$) with magnitude (color scale, not shown) representing the size of the coefficient at that scale and rotation.}
\end{figure}

\begin{figure}[ht!]
\center\includegraphics[width=30pc]{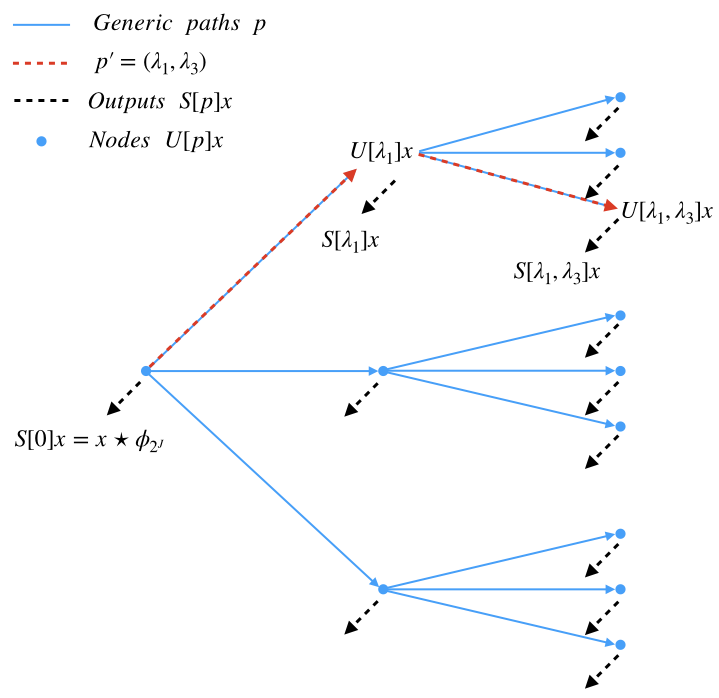}
\caption{\label{fig:mstnet} The MST may be thought of as a convolutional network with fixed weights. The above network could represent for example a 1D MST with $3$ scales, and no rotations (1D case). The network outputs MST coefficients averaged by a Father Wavelet along each path $S[p]x$. Each node of the network is the set of scattering coefficients before being window averaged by the Father Wavelet $U[p]x$. The operator $\tilde{W}$ of Eq.~\ref{eqn:MST} expands the network below a given node at then end of a path $p$.}
\end{figure}

The two-dimensional MST uses a set of convolutional filters which are calculated from a Mother Wavelet $\psi$ by applying a rotation $r$ and scaling by $2^j$:
\begin{equation}
\label{eqn:wavelet}
\psi_{\lambda} = 2^{-2j} \psi (2^{-j} r^{-j} u),
\end{equation}
where $\lambda=2^{-j} r$ and $u$ is the spatial position. Let the wavelet transformation of image $x(u)$ be given by $x \star \psi_\lambda$. Given that the spatial resolution is retained in a wavelet transform, this process can be iterated upon, such that the propagated signal along path $p = (\lambda_1, \lambda_2,\dots,\lambda_m)$ is given by:
\begin{eqnarray}
	U[p]x &= U[\lambda_m] = \cdots U[\lambda_2]U[\lambda_m]x \nonumber \\
	 &= | || x \star \psi_{\lambda_1} | \star \psi_{\lambda_2} | \cdots | \star \psi_{\lambda_m} | \label{eqn:wavelet_of_wavelet}
\end{eqnarray}
where the modulus removes the complex phase from the propagated signal. However, the wavelet coefficients are not invariant to translation, but rather translation covariant. Introducing the Father Wavelet (\textit{i.e.} a spatial window function) $\phi_{2^J}(u)=2^{-2J}\phi(2^{-J}u)$ allows an average pooling operation to be performed by convolution $U[p]x \star \phi_{2^J}(u)$. This operation collapses the spatial dependence of the wavelet coefficients while retaining the dominant amplitude $U[p]$ at each scale. This results in an effective translation invariance assuming that a given translation $c$ is much smaller than the window scale $2^J$. The windowed scattering transformation is thus given by:
\begin{eqnarray}
	S[p]x(u) &=&U[p]x \star \phi_{2^J}(u) \nonumber \\
	&=&| || x \star \psi_{\lambda_1} | \star \psi_{\lambda_2} | \cdots | \star \psi_{\lambda_m} | \star  \phi_{2^J}(u). \label{eqn:windowed_scattering}
\end{eqnarray}
Now, we may define and operator $\tilde{W}$ which acts upon the non-windowed scattering $U[p]x$ producing
\begin{equation}
	\tilde{W} U[p] x = \{S[p]x, U[p + \lambda]x \}_{\lambda \in \mathcal{P}} \label{eqn:MST}.
\end{equation}
$\tilde{W}$ will produce the output scattering coefficient at the current layer for the given path $p$, and will move to the next layer along the path $p+\lambda$ as demonstrated in Fig.~\ref{fig:mstnet}.  With Eqns.~\ref{eqn:wavelet_of_wavelet} and~\ref{eqn:windowed_scattering}, we arrive at a deep scattering convolutional network $\tilde{W}$ (Eq.~\ref{eqn:MST}) with $m$ layers. For 2-D signals (images), the MST coefficients are visualized via log polar plots as depicted in Fig.~\ref{fig:logpolar}. 

In summary the MST forms a nonlinear mapping from an image's spatial features to its scale features. This mapping is Lipschitz continuous to deformation, meaning that small deformations of the image result in small deformations of the Mallat scattering coefficients. Since we will be concerned with discovering morphology parameters of stagnation column images such as helical wavelength, MST provides a convenient basis as compared to, for example, a Fourier transform which is not Lipschitz continuous to deformations. We refer the reader to Refs. \citep{Mallat2012,Bruna2013} for more details regarding the mathematical properties of MST.

\section{Synthetic Model and ML Pipeline}\label{sec:application}

\subsection{Synthetic Double Helix Model}
In order to quantify the morphology of the MagLIF stagnation column, a model with well defined parameters is needed to act as a surrogate for the x-ray self emission diagnostic images. For this purpose we have constructed a synthetic model complete with 11 descriptive model parameters that capture the 2D morphological projection of a fundamentally 3D stagnating plasma along with 6 stochastic parameters to represent the {\textit{natural}} experimental variation and signal noise inherent in the x-ray diagnostics fielded on Z.

Analytically, the synthetic model consists of superimposed radial and axial Gaussians over a pair of $\cos^2$ waves. The model may be specified by the composition of the following functions:
\begin{eqnarray}
\label{eqn:s}
s(z) &=& \theta_6 \cos^2(\theta_7*\theta_3*z+\zeta_5) \nonumber\\ 
&+&\theta_9 \cos^2(\theta_{10}*\theta_3*z+\zeta_6),
\end{eqnarray}
\begin{equation}
\label{eqn:r0}
r_{0,i}(r,z)= (-1)^{1+\delta_{i,2}}\theta_8 + \theta_5 \sin(\theta_3*z + \zeta_4 + \delta_{i,2} \theta_{11}),
\end{equation}
\begin{equation}
\label{eqn:g}
g_i(r,z,r_{0,i}(r,z)) = \frac{1}{\sqrt{2\pi \theta_1}} \exp\Big\{\frac{-(r-r_{0,i}(r,z))^2}{2 \theta_1^2}\Big\},
\end{equation}
\begin{equation}
\label{eqn:ell}
\ell(r,z) = \frac{\zeta_3}{\sqrt{2\pi \theta_2}} \exp\Big\{\frac{-z^2}{2 \theta_2^2}\Big\},
\end{equation}
and
\begin{eqnarray}
\label{eqn:h}
h(r,z) &=& \sum_{i=1}^2\Big[ (1+s(r,z))g_i(r,z,r_{0,i}(r,z))\ell(r,z) \nonumber\\
&\times& (1-\zeta_2 U(0,1)) + \zeta_1U(0,1)\Big],
\end{eqnarray}
where $h(r,z)$ is the final composition used to generate double helix images, $U(0,1)$ is a uniformly distributed random number on $[0,1]$, and $(\theta_i, \zeta_i)$ will be described. 

\begin{figure}[ht]
\center\includegraphics[width=20pc]{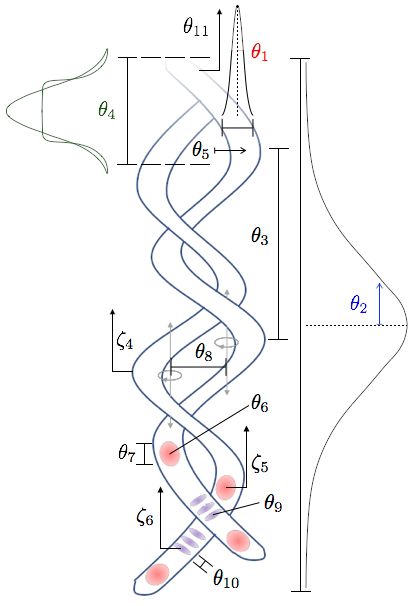}
\caption{\label{fig:syn} Synthetic Stagnation Model  (see Table~\ref{tab:param}).}
\end{figure}

The $\theta_i$ and $\zeta_i$ parameters are depicted in Fig.~\ref{fig:syn} and summarized in Table~\ref{tab:param}. Their interpretations are: $\theta_1$ is the standard deviation of the radial Gaussian,   $\theta_2$ is the standard deviation of the axial Gaussian, $\theta_3$ is the Magneto Rayleigh-Taylor (MRT) wavenumber, $\theta_4$ is the order of the super Gaussian, $\theta_5$ is the helical strand radius, $\theta_6$ is the amplitude of the large-wavelength axial bright spot, $\theta_7$ is the mode of the large-wavelength axial bright spot, $\theta_8$ is the strand separation, $\theta_9$ is the amplitude of the small-wavelength axial bright spot, $\theta_{10}$ is the mode of the small-wavelength axial bright spot and $\theta_{11}$ is the strand phase; $\zeta_1$ is the background noise, $\zeta_2$ is the signal noise, $\zeta_3$ is the amplitude of the signal, $\zeta_4$ is the radial perturbation phase shift, $\zeta_5$ is the is the phase shift of the large-wavelength axial bright spot and $\zeta_6$ is the phase shift of the small-wavelength axial bright spot.

\begin{figure*}[ht!]
\includegraphics[width=\columnwidth]{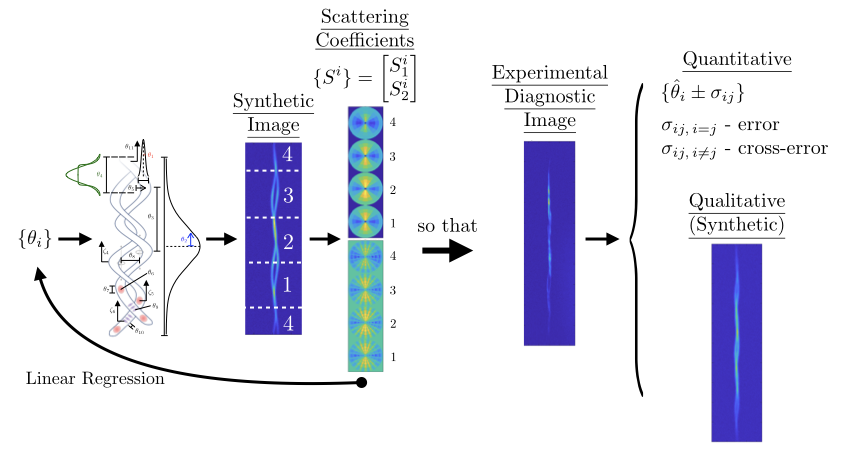}
\caption{\label{fig:mstr}  The MST regression pipeline for morphology characterization of experimental stagnation images.}
\end{figure*}

Our ultimate goal will be to create a machine learning pipeline which can take as input an image of a plasma stagnation column and output a set $\{\theta_i\}_{i=1}^{11}$ characterizing the morphology of the column along with an estimate of the uncertainty in our output. This will be achieved by creating a set of synthetic images from Eq.~\ref{eqn:h} using a large set of randomly chosen $(\theta_i,\zeta_i)$, computing the MST, and performing a regression from MST coefficients to $\theta_i$. We reserve the details of how error estimates are obtained for Sec.~\ref{sec:reg}. Note the absence of $\zeta_i$ in our output as those are meant to represent unimportant transformations, such as rotating the viewing angle, which does not alter the fundamental morphology. Our pipeline is summarized in Fig.~\ref{fig:mstr}. However, before considering the full regression problem, it is instructive to first explore the ability of the MST to distinguish between double helix images with very different appearance. To this end, we begin not with the full regression problem, but rather with a classification problem in next section.

\begin{table}
 \centering
	    \caption{Synthetic model $\theta_i$ and stochastic $\zeta_i$ parameters (see Fig.~\ref{fig:syn}).}
	    \bigskip

\begin{tabular}{l}
\textbf{Model Parameters}\\
\hline
$\theta_1$ = thickness\\
$\theta_2$ = length\\
$\theta_3$ = helical wavenumber\\
$\theta_4$ = order of axial super Gaussian\\
$\theta_5$ = amplitude of radial perturbations\\
$\theta_6$ = amplitude of large-wavelength axial brightness\\
perturbations\\
$\theta_7$ = mode number of large-wavelength axial brightness\\
perturbations\\
$\theta_8$ = strand separation\\
$\theta_9$ = amplitude of small-wavelength axial brightness\\
perturbations\\
$\theta_{10}$ = mode number of small-wavelength axial brightness\\
perturbations\\
$\theta_{11}$ = relative strand phase\\
\\
\textbf{Stochastic Parameters}\\
\hline
$\zeta_1$ = background noise\\
$\zeta_2$ = signal noise\\
$\zeta_3$ = amplitude of signal\\
$\zeta_4$ = radial perturbation phase shift\\
$\zeta_5$ = large-wavelength axial brightness perturbations phase\\shift\\
$\zeta_6$ = small-wavelength axial brightness perturbations\\ phase shift
\end{tabular}

\label{tab:param}
\end{table}

\subsection{Classification Model}\label{sec:class}
\begin{figure*}[ht]
\includegraphics[width=\columnwidth]{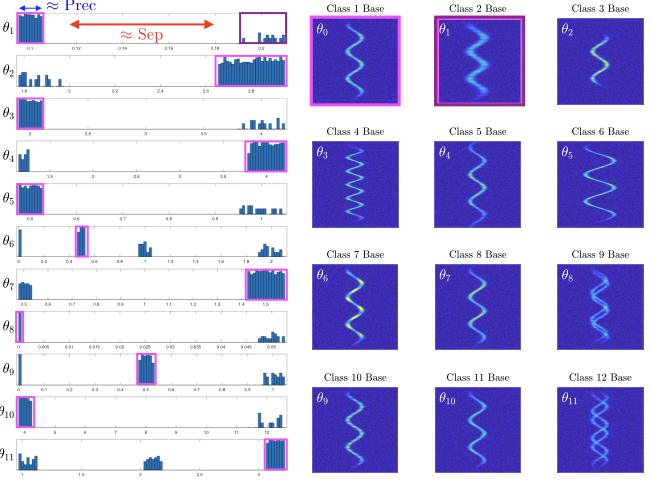}
\caption{\label{fig:classes} Classification Training Set construction. Parameter distributions are shown at left while the base classes are represented on the right.  Shown is a class separation as the red line labeled ``Sep'', and a class precision as the blue line labeled ``Prec''.}
\end{figure*}
Studying the ability of the MST to distinguish between different classes of helical morphology will provide reassurance that the regression problem is well-posed. Additionally, it provides access to more easily interpretable results (e.g. classification accuracy as opposed to $R^2$). By considering the classification problem, we are also able to closely follow the approach using MST for MNIST handwritten digit recognition in \citet{Bruna2013}. Indeed, a majority of the design decisions discussed below including image gridding, affine space dimensionality, and scale resolution are inherited from \citet{Bruna2013}.

We approach the problem by synthesizing 12 stagnation image classes -- 11 distinct parameter constrained classes constructed from systematic modifications to the synthetic model parameter distributions from a single base class. Each of the distinct parameter classes has a definitive associated synthetic model parameter. For a given parameter class, the distribution of its associated synthetic model parameter is translated some separation from its corresponding base class distribution. This process is repeated for each of the 11 distinct parameter classes (see Fig.~\ref{fig:classes}). For the classification problem, we generate 340 images. We use $50\%$ of this data set as the training set to train an affine classifier, while the remaining $50\%$ is separated out as the test set to be used for characterizing the trained classifier.

\begin{figure*}[ht]
\includegraphics[width=\columnwidth]{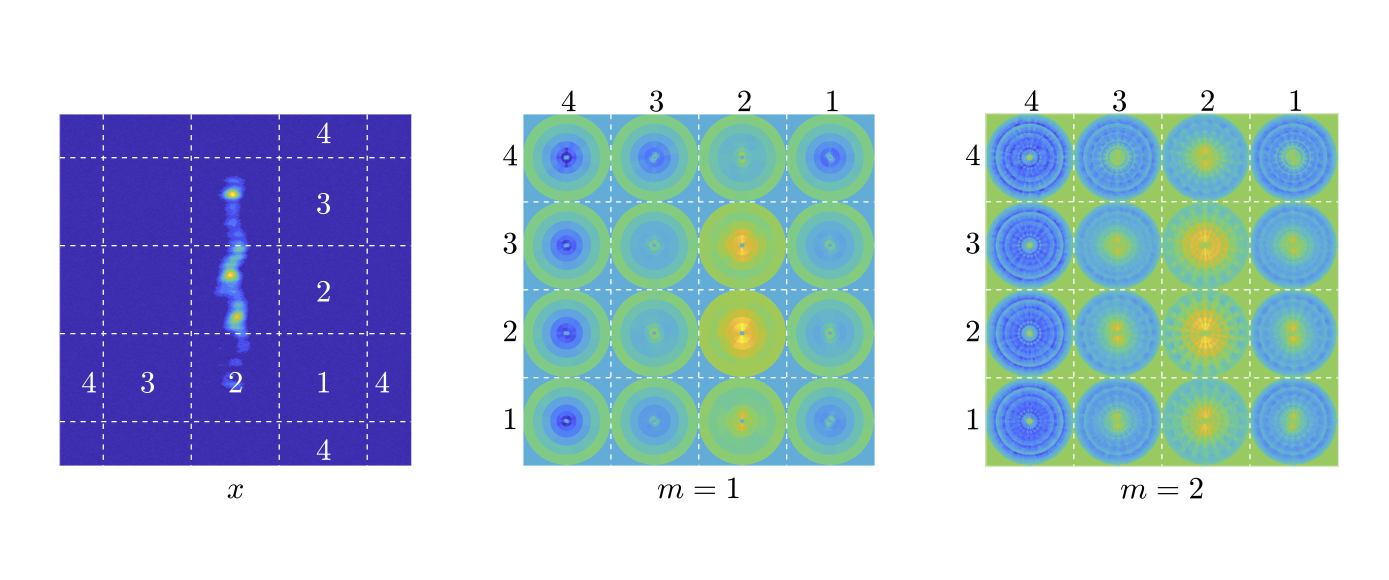}
\caption{\label{fig:grid} Gridded MST: (a) image, (b) first and (c) second order MST coefficients.}
\end{figure*}
We now discuss how features were engineered from this dataset using the MST. First, the reader may note from Eq.~\ref{eqn:windowed_scattering} that we must evaluate the scattering coefficients at points $u$ in our image. Now, due to the assumption that the statistics given by the MST are stationary, that is spatially invariant, below the Father Wavelet window size; if we were to evaluate $S[p]x(u)$ at all points $u$, one would obtain very redundant information. As a result, it is wise to subsample $u$. This is achieved by translating the spatial window by intervals of $2^J$ such that $G_\#=N2^{-J}$, where $N$ is the symmetric pixel count and $G_\#$ is symmetric grid number. This subsampling forces each image to be segmented into a $G_\#\times G_\#$-grid \citep{Bruna2013}. We work with images of pixel size $512\times512$, and set $J=7$ giving $G_\#=4$.  We now have a design parameter to choose, $J$, which determines the size of the sub-image, $2^J \times 2^J$, over which the transform will be calculated. This was chosen based on the position, size and characteristics of our double helix (see Fig.~\ref{fig:grid}) and was found to give good classification accuracy as discussed later. We note however, that a more rigorous procedure to select $J$ would be to select the $J$ which gives maximum cross validated classification accuracy as in \citet{Bruna2013}.  With this being said, our eyes are very good at recognizing the dynamical space scale of the physics (see Sec.~\ref{sec:wwt} for a detailed description of this scale).  The size of the Father Wavelet, $2^J$, should be of this scale.  If the size is too small, the MST will be not be calculated over the largest area possible and will therefore have more noise and not contain as much statistical information.  If the size is too large, the assumption of stationarity will be violated leading to a blurring of the statistics and a resulting loss of information.  It is therefore expected that there will be an optimal size that could be determined by the aforementioned $J$ cross validation optimization.

An added benefit to gridding the images, is data reduction via patch selection. From Fig.~\ref{fig:grid}(a) it is apparent that most of the image is background noise. This is echoed in the MST coefficient space. Since our double helix is confined to column $2$, essentially all of the unique information is contained within the MST coefficients evaluated on the four patches in column $2$, so that the other columns may be dropped.

Before computing the MST on our gridded image, we must apply boundary conditions for the convolution. There are many reasonable choices,  such as periodic, zero-padded, and mirrored. We chose to use a mirror boundary condition, but found very little impact to our classification accuracy as compared to periodic.

The final step in engineering features for a classification algorithm is to perform an appropriate scaling of the input features. This is a common practice in statistical learning, and many different scaling transformations and dimensionality reduction methods are reasonable. Here, we apply a $\log_{10}$ scaling to our scattering coefficients and model parameters (with the exception of $\theta_{11}$ which is a phase shift) used in training the classifier. This choice was made to decrease the dynamic range of the MST coefficients, since before the transformation the MST coefficients were dominated by only a few coefficients.

Finally, we may apply the classification algorithm. Following \citet{Bruna2013} we apply a classifier based on an affine space model with the approximate affine space determined by principal component analysis of each class. To be specific, let $SX_k$ denote the set of MST coefficients for all of our images belonging to class $k$. $SX_k$ can be organized into a $N_{i,k}\times P$ matrix where $N_{i,k}$ is the number of images available for class $k$ and $P$ is the number of scattering coefficients (\textit{i.e.} the coefficients have been stacked into a vector of length $P$). The columns of $\Delta_k$ may be transformed to have zero mean for each of the $P$ coefficients $\Delta_k = SX_k - \mathbb{E}(SX_k)$. We may then perform principal component analysis on $\Delta_k$ by finding the eigenvectors $\{\mathbf{U}_{j,k}\}_{j=1}^P$ and corresponding eigenvalues $\{\Lambda_j\}_{j=1}^P$ of the covariance matrix $\Delta_k^T\Delta_k$. Taking $\mathbf{U}_{j,k}$ to be ordered such that $\Lambda_j > \Lambda_{j+1}$, we keep only the first $d \ll P$ principal vectors $\{\mathbf{U}_{j,k}\}_{j=1}^d$. Letting $\mathbf{V}_k = \text{span}(\{\mathbf{U}_{j,k}\}_{j=1}^d)$,  we may construct the affine approximation space for class $k$
\begin{equation}
\label{eqn:affinespace}
\mathbf{A}_k = \mathbb{E}(SX_k) + \mathbf{V}_k.
\end{equation}
Finally, for a new image with scattering coefficients $Sx$, the class assigned to the image is given by 
\begin{equation}
\hat{k}(x) = \underset{k}{\operatorname{argmin}} || Sx - P_{\mathbf{A}_k}(Sx)||.
\end{equation}

\begin{figure}[ht]
\center\includegraphics[width=20pc]{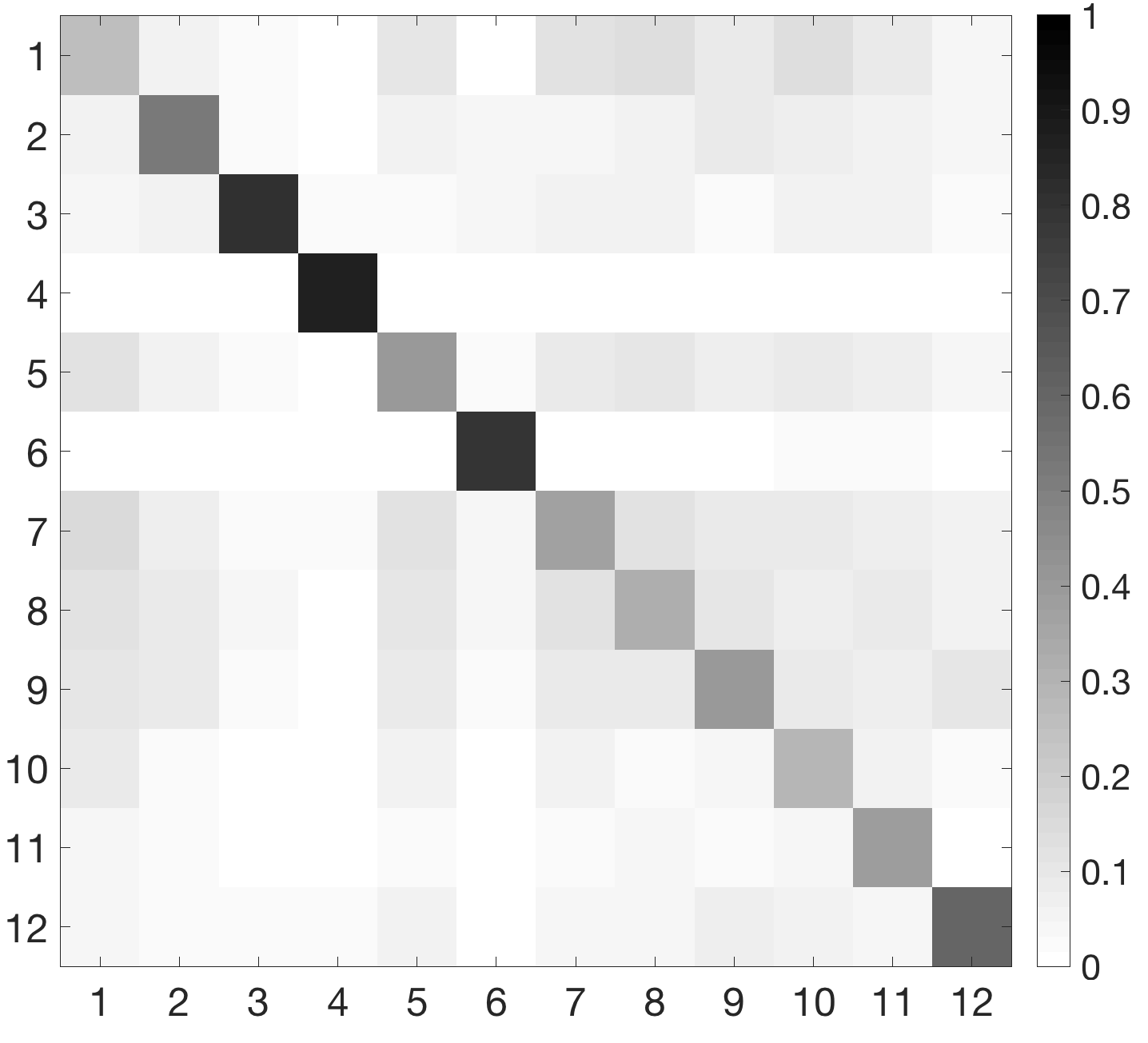}
\caption{\label{fig:confusion} The matrix $\Omega$ defined by Eq.~\ref{eqn:sepdec} which demonstrates that the constructed double helix classes are well separated in the MST space.}
\end{figure}

%We keep only the first principal component in the affine approximation. ($d=1$).%
In order to evaluate how effectively the classes are separated one may define the ratio of the affine approximation error in a class $i$ to the error in class $j$. 

\begin{equation}
\label{eqn:sep}
R^2_{ij}= \frac{E(|| SX_i - P_{\mathbf{A_j}}(SX_i)||^2)}{E(|| SX_i - P_{\mathbf{A_i}}(SX_i)||^2)}.
\end{equation}
Note that if the classes are well separated, then $R_{i,j}^2$ will be very large for $i\neq j$, while $R_{i,i}^2=1$. It thus makes sense to define the matrix
\begin{equation}
\label{eqn:sepdec}
\Omega_{i,j} =N_j e^{-R^2_{ij}},
\end{equation}
where $N_j$ is a column-wise normalization ensuring that each column of $\Omega$ sums to $1$. Fig.~\ref{fig:confusion} shows the matrix $\Omega$ for our case demonstrating good class separation as indicated by the fact that the matrix is strongly diagonal.  The off-diagonal elements are indicative of overlap among the the tails of the class distributions. The chance of miss-classification is extremely small ($<0.1\%$), and the average class precision is $0.0016$ while the average class separation is $8.14$.  Here we have used the definitions of class separation and precision given in \citet{Bruna2013}.  Note that the separation is just the average of the separation matrix given in Eq.~\ref{eqn:sep}.  The geometric meaning of the separation and precision are shown in Fig.~\ref{fig:classes}.

\subsection{Regression Model}\label{sec:reg}
We now consider the full regression problem as highlighted in Fig.~\ref{fig:mstr}. The early stages of the pipeline are not significantly altered. For example, we still use $\log_{10}$ scaled MST coefficients. However, rather than generating distinct classes, we generate images by selecting $\theta_i$ values from a range which will contain all of the classes discussed in the classification problem. Specifically,  image realizations are produced, using the synthetic model, from a random sampling of the log-uniformly distributed model parameters. The statistical properties of these distributions are determined by visually confirming that helices produced encompass what is reasonable to expect from experiment. Additionally, most of the quantities we wish to learn from the helical images (\textit{i.e.} the $\theta_i$'s) are non-negative. As a result, we chose to $\log_{10}$ scale all of the $\theta_i$ values except for the strand phase $\theta_{11}$. The only other difference is the replacement of the affine classifier with a linear regression method which we discuss below. 

\begin{figure*}[ht]
\includegraphics[width=\columnwidth]{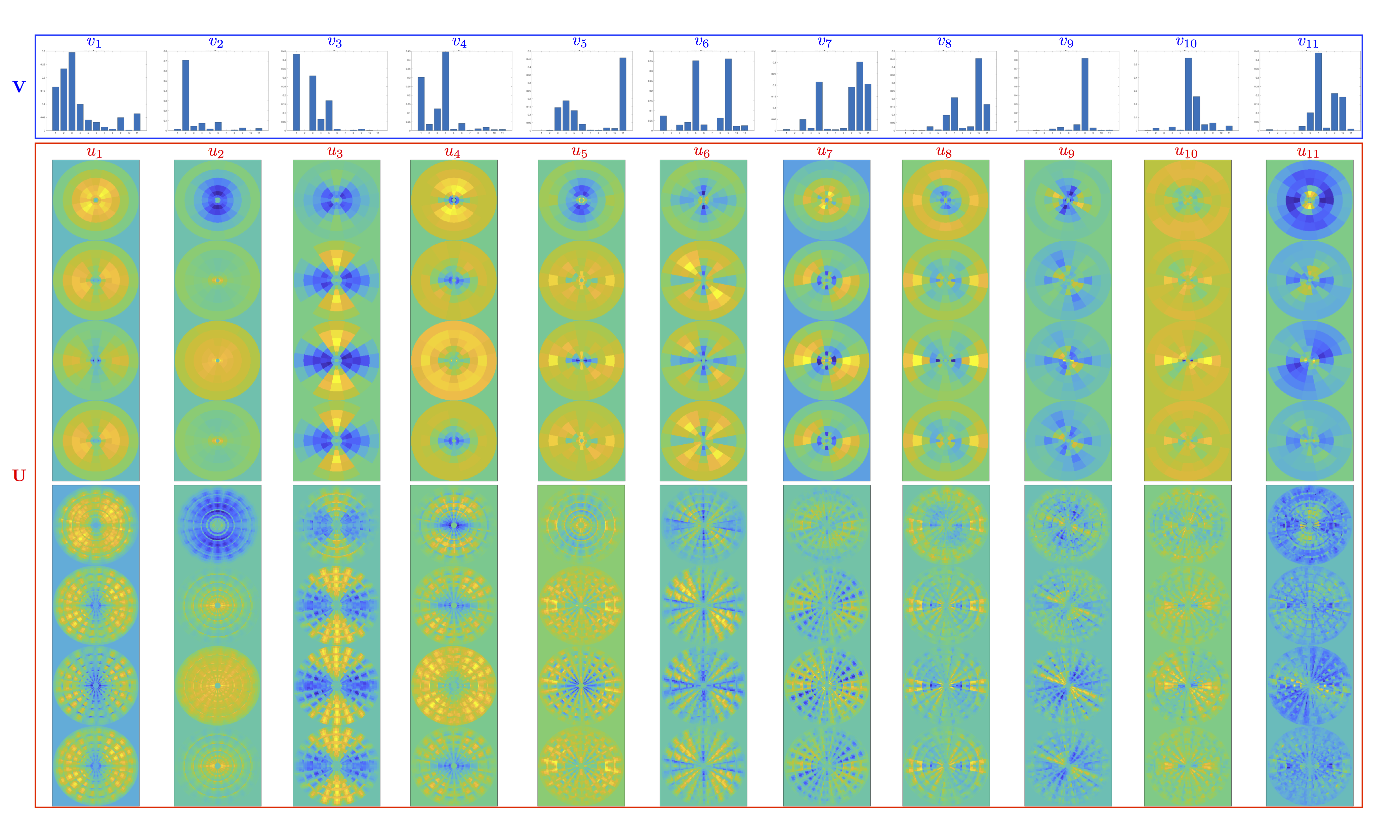}
\caption{\label{fig:pcv} Linearization of the model parameter-scattering coefficient space. Orthogonal basis vectors, $\mathbf{V}$ and $\mathbf{U}$, mapping the model parameters (top panel, $\mathbf{V}$) and scattering coefficients (bottom panel, $\mathbf{U}$) into the linearized space.  For the scattering coefficients, the first order is shown in the top row and the second order in the bottom row.}
\end{figure*}

Before conducting a linear regression from ($\log_{10}$ scaled) MST coefficients to (scaled) helical parameters, we standard normal scale $\theta$ and $\mathbf{S}$. We will henceforth refer to the transformed quantities as $\tilde \theta$ and $\mathbf{\tilde S}$. We can then convert to a basis in which the cross-covariance between the transformed scattering coefficients and model parameters from our training set, CCOV$( \mathbf{\tilde \theta}, \mathbf{\tilde S}) = \mathbf{\tilde \theta}^T\mathbf{\tilde S}/(N-1)$ is diagonal. Here, $N$ is the number of training samples used to construct the cross-covariance. To do this, we take the singular value decomposition (SVD) of the cross-covariance. The SVD factors the cross-covariance matrix into a set of transformation matrices $\mathbf{U}$ and $\mathbf{V}$ bound by a diagonal matrix $\mathbf{\Sigma}$ containing a set of singular values
\begin{equation}
\label{eqn:svd}
\mathbf{U \Sigma V}^T = \frac{\mathbf{\tilde \theta}^T\mathbf{\tilde S}}{N-1}.
\end{equation}
This set of transformation matrices provide a set of orthogonal bases vectors along which $\tilde{\theta}$ and $\tilde{\mathbf{S}}$ are most strongly linearly correlated ordered from strongest to weakest correlation (see Figure~\ref{fig:pcv}). The model parameters $\tilde{\theta}$ and scattering coefficients $\tilde{\mathbf{S}}$ are rotated into the linearized space such that $\mathbf{Y}=\mathbf{\tilde \theta U}$ and $\mathbf{X}=\mathbf{\tilde S V}$ define the linearized variables, respectively.

Regressing the linearized scattering coefficients $\mathbf{X}$ back onto the linearized $\mathbf{Y}$ is accomplished using multidimensional linear regression as shown in Eq.~\ref{eqn:reg}, where $\mathbf{m}$ is the map from $\mathbf{X}$ to $\mathbf{Y}$ (e.g. "slope"), $\mathbf{b}$ is the bias (e.g intercept) and $\mathbf{\epsilon}$ is the error term 
\begin{equation}
\label{eqn:reg}
Y_j = b_j + \sum_{i=1}^{p}X_i m_{ij} + \epsilon_j.
\end{equation}
where $\epsilon =\mathcal{N}(0,\mathbf{\Lambda})$ is assumed to be a zero mean normal random variable with covariance matrix $\mathbf{\Lambda}$. Writing Eq.~\ref{eqn:reg} in matrix notation, the bias is absorbed into the slope such that $\mathbf{Y}=\mathbf{XM}+\mathbf{\epsilon}$. 

Note that Eq.~\ref{eqn:reg} implies that the prediction for a new input $\mathbf{X}$ is $\mathbf{Y}_\text{pred} =\mathbf{ \overline{Y}} = \mathbf{XM}$ since $\overline{\mathbf{\epsilon}}=0$. Importantly, would also be able to characterize the uncertainty in our prediction if we had an estimate of $\mathbf{\Lambda}$. In order to estimate $\mathbf{M}$ and $\mathbf{\Lambda}$, note that Eq.~\ref{eqn:reg} specifies a likelihood function 
\begin{equation}
\begin{split}
P(\{\mathbf{Y}_i\}|\mathbf{M},\mathbf{\Lambda},\{\mathbf{X}_i\}) = \prod_{i=1}^{N} \frac{1}{\sqrt{(2\pi)^k|\mathbf{\Lambda}|}}\\
\times e^{-\frac{(\mathbf{Y}_i-\mathbf{X}_i\mathbf{M})^T\mathbf{\Lambda}^{-1}(\mathbf{Y}_i-\mathbf{X}_i\mathbf{M})}{2}},
\end{split}
\end{equation}
where the training data are assumed i.i.d. and $k$ is the dimensionality of our output space (here $k=11$ since there are $11$ theta parameters we wish to regress to).
A maximum likelihood estimate of the coefficients of the map matrix $\mathbf{M}$ and error covariance matrix $\mathbf{\Lambda}$ are determined by finding their values which maximize the likelihood function over our training data. Equivalently, since the logarithm is monotonic, we may maximize the log-likelihood $\mathcal{L}$. The solution is derived in many statistics and machine learning textbooks (see \textit{e.g.} \citet{bishop}) and is given by
\begin{equation}
\label{eqn:mleM}
\mathbf{M}_\text{MLE} = (\mathbf{X}^T\mathbf{X})^{-1}\mathbf{X}^T\mathbf{Y},
\end{equation}
which is the typical ordinary least squares solution where $\mathbf{X}_i$($\mathbf{Y}_i$) have been stacked to create $\mathbf{X}$($\mathbf{Y}$) and the error covariance matrix is
\begin{equation}
\label{eqn:mleL}
\mathbf{\Lambda}_\text{MLE}= \frac{1}{N} \sum_{i=1}^N(\mathbf{Y}_i-\mathbf{X}_i\mathbf{M}_\text{MLE})^T(\mathbf{Y}_i-\mathbf{X}_i\mathbf{M}_\text{MLE}),
\end{equation}
which is just the estimate of the population covariance matrix of the difference $(\mathbf{Y}-\mathbf{Y}_\text{pred})$.

\begin{figure*}[ht]
\includegraphics[width=\columnwidth]{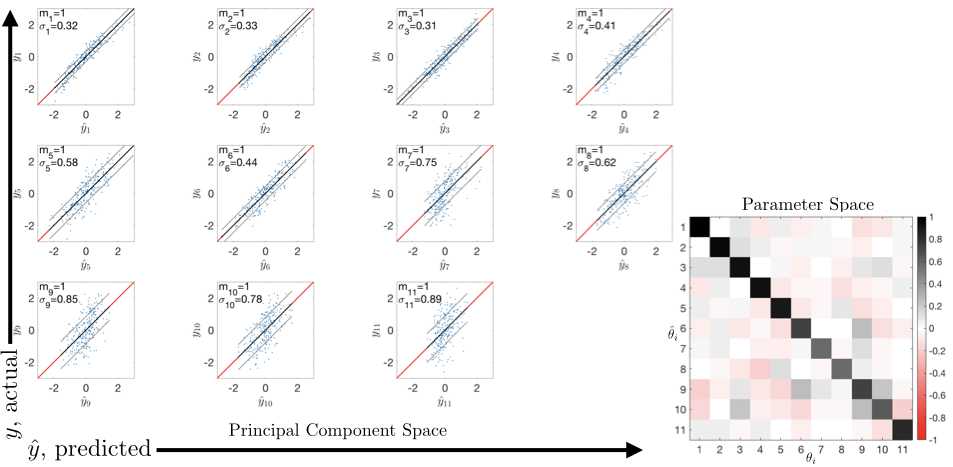}
\caption{\label{fig:mstr_performance} Here we visualize the MST regressor performance. The scatter plots show predicted vs. actual principal components. The first several demonstrate good agreement, while the later components show worse agreement. This may be indicative of nonlinearity which the linear regression cannot explain, or it may be variance caused by the unexplained $\zeta$ parameters. The correlation plot at right shows that predictions in the original parameter space are essentially uncorrelated.}
\end{figure*}
For a new image, we can now estimate a set of values $\theta$ along with an estimate of the uncertainty on theta according to the following algorithm:
\begin{enumerate}
  \item Compute the first and second order scattering coefficients of the image on a 4x4 grid (see Fig.~\ref{fig:grid}).
  \item Discard all but the second column from the grid for each of the 2 sets of coefficients.
  \item Compute $\log_{10}$ of the scattering coefficients and flatten into a vector to get $\mathbf{S}$.
  \item Standard normal scale using the mean and standard deviation estimated on the training set to get $\mathbf{\tilde S}$.
  \item Project onto principal components to get $\mathbf{X} = \mathbf{\tilde S} \mathbf{V}$.
  \item Compute $\mathbf{Y}_\text{pred} = \mathbf{X}\mathbf{M}_\text{MLE}$.
  \item Create a set of values consistent to within the error term 
  \begin{equation*}
  \{\mathbf{Y}_\text{pred,i}\}_{i=1}^{N_\text{resamp}} = \mathbf{Y}_\text{pred} + \{\mathcal{N}_i(0,\mathbf{\Lambda}_\text{MLE})\}_{i=1}^{N_\text{resamp}}.
  \end{equation*}
  \item Compute $\{\tilde \theta_i\} = \{\mathbf{Y}_{\text{pred},i}\mathbf{U}^{-1}\}$.
  \item Compute $\{\theta_i\}$ by inverting standard normal scaling of $\tilde \theta$ using the mean and standard deviations of $\tilde \theta$ computed from the training set and then invert the $\log_{10}$ scaling performed on all but the last component of $\theta$.
  \item We now have an estimate of the distribution of $\theta$ consistent with the original image. We may report the prediction and error as means and standard deviations, or as percentiles (\textit{e.g.} report the $50^{th}$ percentile as the prediction and the $2.5$-percentile and $97.5$-percentile as lower and upper bounds).
\end{enumerate}
In our case, inverting the transformations leads to an asymmetric distribution of $\theta$ values consistent with the original image, so here we will report the $95\%$ confidence interval and the mode of the distribution rather than mean and standard deviation for any predictions.

Before moving on to discuss results, we note that the cross-covariance matrix computes the linear correlation between the quantities $\tilde{\theta}$ and $\tilde{\mathbf{S}}$. As a result, any nonlinear relationships between $\tilde{\theta}$ and $\tilde{\mathbf{S}}$ will not be recoverable upon linear regression. An exploration of the possibility of a nonlinear regression between $\mathbf{S}$ and $\theta$ is reserved for future work. For now, we focus on performance of the linear regression, as our results already indicate some ability to distinguish morphology among different experimental configurations. In the next section we discuss results and implications of the application of our methods to comparing morphology of different stagnation columns.

\section{Results and Discussion}\label{sec:results}
There are two primary cases of interest for applying our method. The first is to be able to quantitatively compare experimental data to simulation. The second is to be able to compare morphology between different experiments. In doing so, we will be able to make statistically sound inferences about discrepancies in morphology. By providing this capability, we hope the method will provide physical insight into mechanisms causing unexpected discrepancies in morphology in the two cases mentioned. To this end, we conduct some initial studies which show how the method could be used in general.

\subsection{Simulation-to-Experiment Comparison}
\begin{figure}[ht]
\includegraphics[width=\columnwidth]{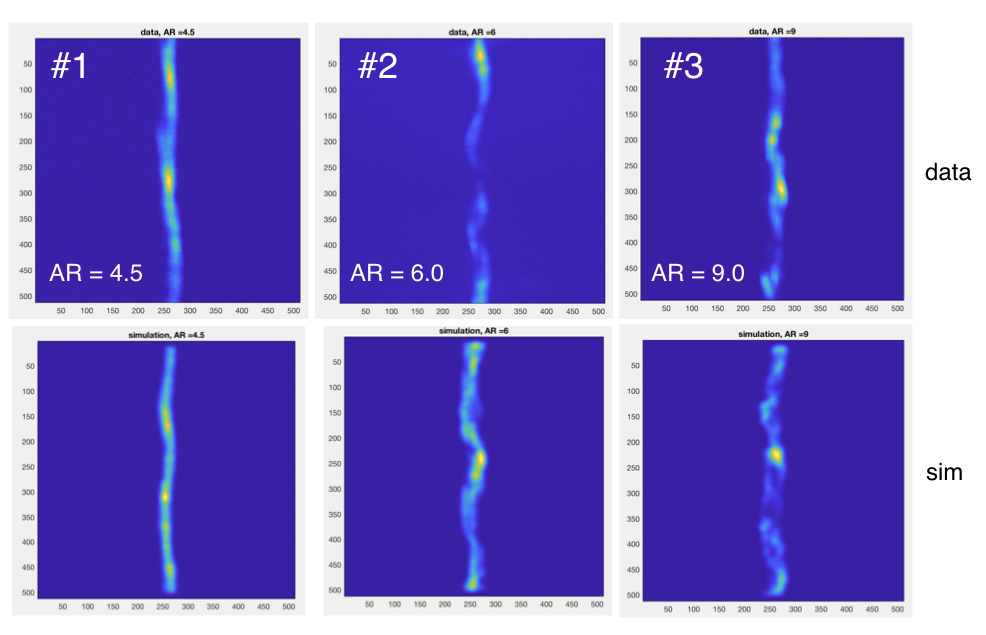}
\caption{\label{fig:simtoexp} Experimental vs 3D Gorgon rad-MHD simulated data for several different aspect ratios.}
\end{figure}

\begin{figure}[ht]
\center\includegraphics[width=20pc]{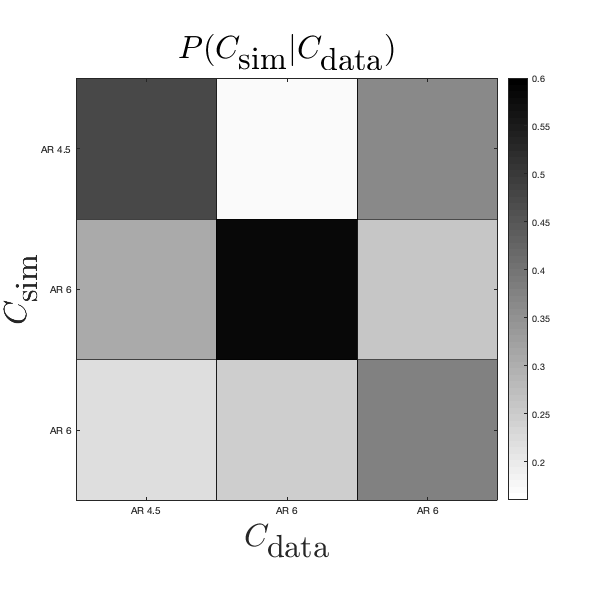}
\caption{\label{fig:sim_vs_exp} Probability of Classification, $\Omega_{i,j}$: probability of classifying $C_\text{sim}$ given that it is $C_\text{data}$.}
\end{figure}

Figure~\ref{fig:simtoexp} shows a comparison of simulated (with the program Gorgon) and experimental data at several different liner aspect ratios.  The MST coefficients, $x$, for the AR$6$ case are shown in the left two columns of Fig.~\ref{fig:sim_vs_exp}. One immediately notices the similarity of the MST coefficients between the two cases. However, there is some nontrivial background present in the experimental data, which our approach lends itself to projecting out. Specifically, if we take the first principal component of the covariance between simulation and experiment, we find the center column of Fig.~\ref{fig:sim_vs_exp}  -- the background, $B$. After projecting out this background component from the experimental data (the right two columns of Fig.~\ref{fig:sim_vs_exp}, $\tilde{x} \equiv x-\left<x,B\right>$), we can observe similarities and differences between the simulation and experimental morphologies by comparing the overall separation of the scattering coefficients, computed as pairwise Euclidean distances, $\sigma_{i,j}$ among the $3$ cases. We can then visualize how well separated they are by plotting
\begin{equation*}
\Omega_{i,j} = N_j e^{-\sigma^2_{ij}},
\end{equation*}
which behaves similarly to $\Omega_{i,j}$ defined in Eq.~\ref{eqn:sepdec}. This quantity is shown in Fig.~\ref{fig:sim_vs_exp} which demonstrates that the 3D Gorgon simulations are generally close in MST space to the corresponding experiment, with the AR$4.5$ simulation showing some pairwise similarity to the AR$9$ experiment. 

\begin{figure*}[ht]
\center\includegraphics[width=30pc]{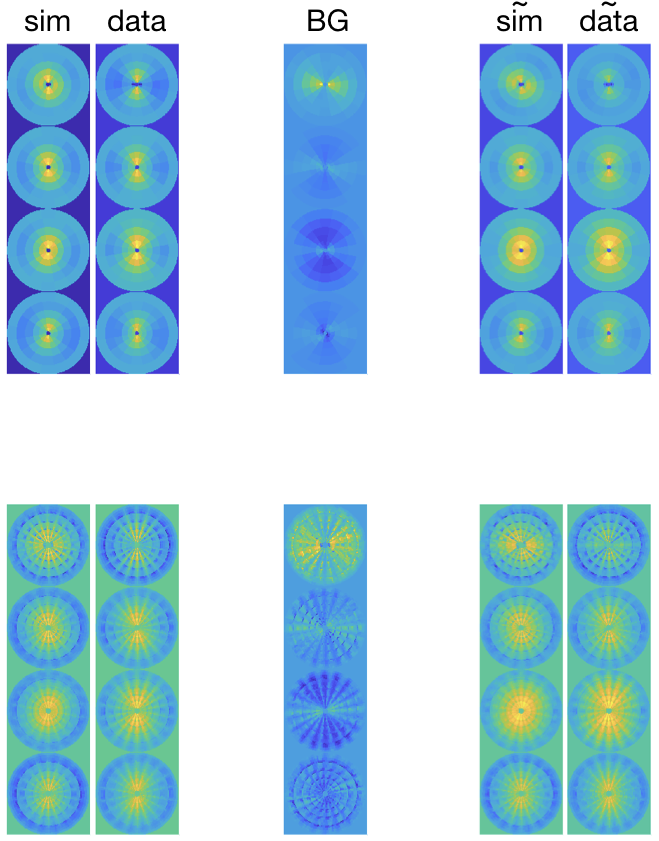}
\caption{\label{fig:bs} Background Subtraction. The top row are the first order MST coefficients, and the second row are the second order MST coefficients.  The two columns on the left are before the background is projected out, the center column is the background derived from the first principal component of the covariance between simulation and experiment, and the right two columns are after the background is projected out.  This is for the AR$6$ case.}
\end{figure*}

\subsection{Experiment-to-Experiment Comparison}
\begin{figure}[ht]
\center\includegraphics[width=25pc]{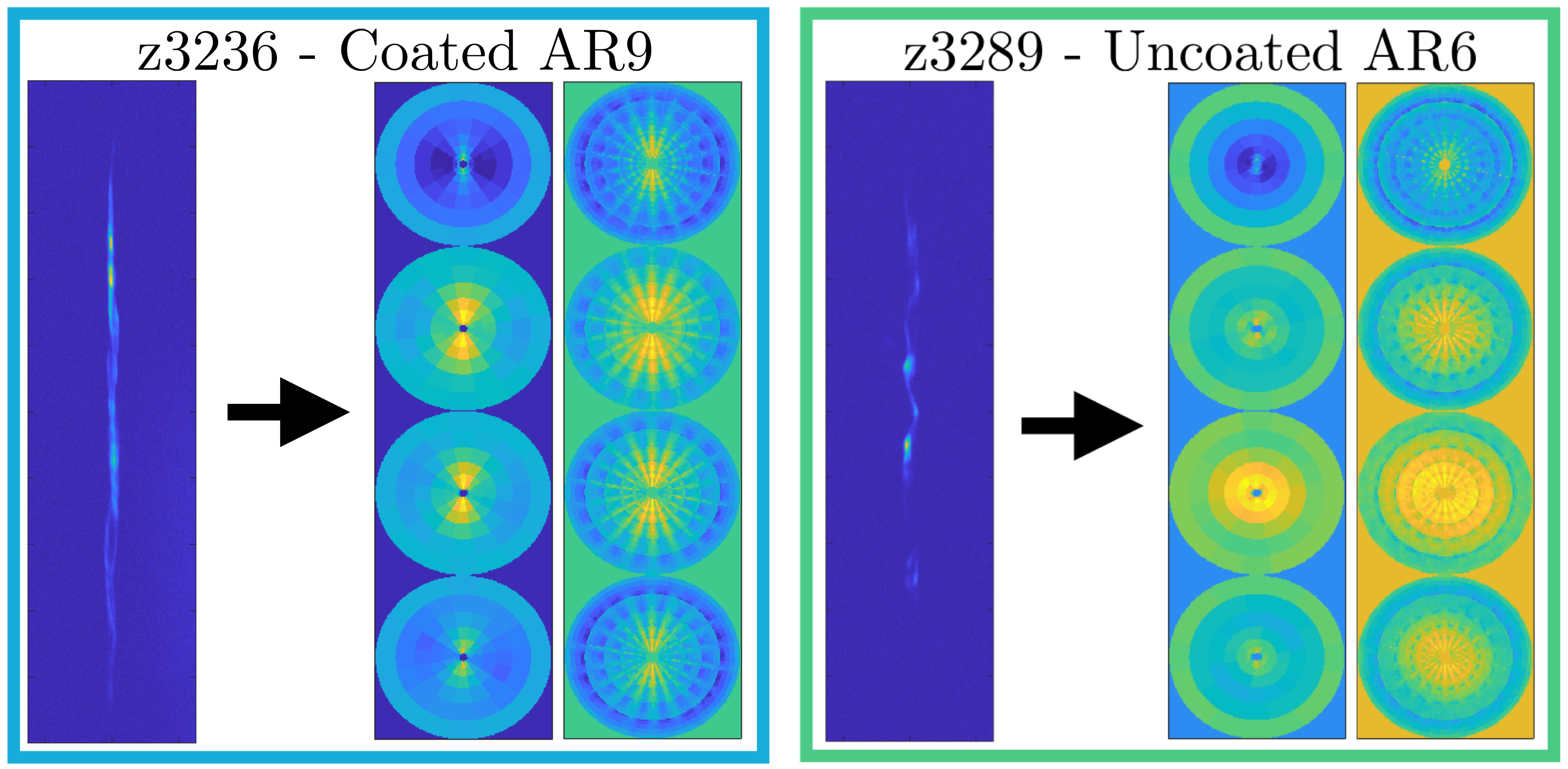}
\caption{\label{fig:coat_vs_uncoat_mst} Comparison of two experiments using the MST.  To the left is shot z3236 with a coated AR9 liner.  To the right is shot z3289 with an uncoated AR6 liner.  Shown, for both cases, are the original stagnation image on the left and the MST on the right (both first and second order coefficients).}
\end{figure}

\begin{figure*}[ht]
\includegraphics[width=\columnwidth]{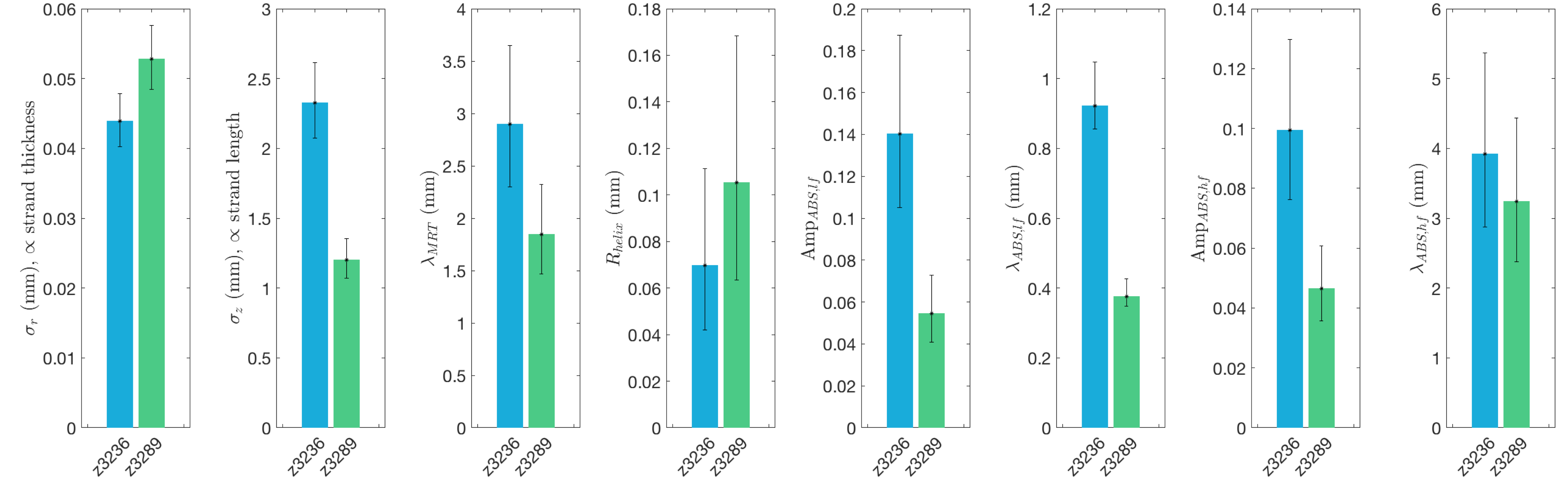}
\caption{\label{fig:coat_vs_uncoat_fit} Regressed parameters of the synthetic helical model for the two experiments shown in Fig.~\ref{fig:coat_vs_uncoat_mst}.  The values for shot z3236 are shown in blue on the left, and for shot z3289 in green on the right. Plotted are modes with error bars showing the 95\% confidence interval.  The error bars are asymmetric because this is not in log space. The parameters are (from left to right): strand thickness (mm), strand length (mm), helical MRT wavelength (mm), radius of the helix (mm), amplitude of the low frequency axial brightness perturbations, wavelength of the low frequency axial brightness perturbations, amplitude of the high frequency axial brightness perturbations, and wavelength of the high frequency axial brightness perturbations.}
\end{figure*}

Finally, we finish with a discussion of differentiating morphology between experiments. Figure~\ref{fig:coat_vs_uncoat_mst} shows the experimental plasma stagnation columns for two different liner designs and their first and second order MST coefficients. To the left is experiment z3236 which utilized a dielectric coated AR9 target, while to the right is experiment z3289 which had an uncoated AR6 liner.  There obvious differences between the MST coefficients for the two cases; but what is different?  To answer this question, we applied the regression derived in the previous section.  The results are shown in Fig.~\ref{fig:coat_vs_uncoat_fit}.  The estimates of selected parameters of the synthetic helical model, the $\theta$'s, along with their uncertainties, are plotted for the two cases side-by side.  For parameters such as the radius of the helix and the wavelength of the high frequency axial brightness perturbations; there is no difference.  For other parameters such as the strand thickness, and helical MRT wavelength; there are modest differences.  For yet other parameters such as the strand length, the amplitude of the low frequency axial brightness perturbations, the wavelength of the low frequency axial brightness perturbations, and the amplitude of the high frequency axial brightness perturbations; there are significant differences.

\section{Physical Interpretation of MST}\label{sec:interpretation}
The MST was originally constructed to have the properties of both deformational continuity (Lipschitz continuity), as well as limited translational invariance (local stationarity).  This combined the best properties of both the wavelet transform and Fourier transform.  It was then found to be unitary (energy conserving), and a straightforward way was found to build in any group symmetry.  When it was used as a coordinate transformation, in the context of modern data science, it was found to have excellent performance in classification and regression of images and signals.  The signals and images seemed to be sparse in this basis.  Little information, if any, was found in the coefficients of greater than second order, but large additional value was found in the second order coefficients compared to the first order coefficients.  The question is why?  This section explores two ways that the transformation can be directly related to the causal dynamics of physical systems.  The first is from a classical perspective, and the second is from a field theoretical perspective.  In addition to explaining the performance of the MST, this formalism will also allow direct interpretation and use of the scattering coefficients.  The conclusion will be that the first order scattering coefficients give the current state of the system, and the second order coefficient give how the system will evolve.

\subsection{Classical:  Manifold Safe Wigner-Weyl Transformation}\label{sec:wwt}
\begin{figure}[ht]
\center\includegraphics[width=25pc]{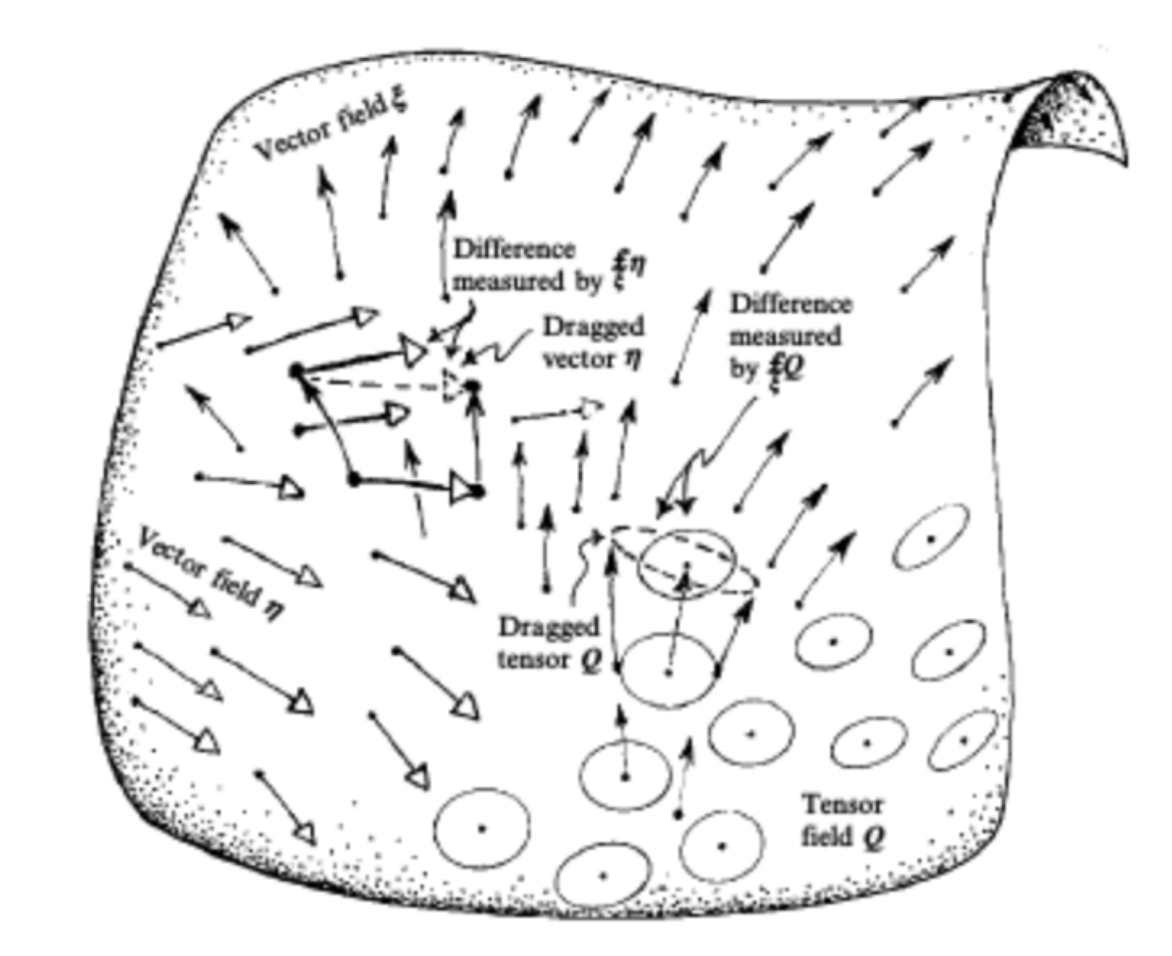}
\caption{\label{fig:advect} Advection of a generalized vector field on a manifold.}
\end{figure}

We will start with the classical perspective of what the MST is.  We first need to observe that causal dynamics can be expressed in a Generalized Liouville Equation using the notation of coordinate free exterior calculus,
\begin{equation}
\label{eqn:gle}
\frac{\partial \rho^{(N)}}{\partial t} + \mathscr{L}_{u^{(N)}} \rho^{(N)} = 0,
\end{equation}
where
\begin{equation}
\rho^{(n)} \equiv f_n \tau^{(n)}
\end{equation}
is the $n$-particle distributions form, $f_n$ is the $n$-particle distribution function, $\mathscr{L}$ is the Lie derivative,
\begin{equation}
\tau^{(n)} \equiv \prod_{i=1}^{n}{\wedge \, \omega_i},
\end{equation}
$\omega=\text{d}\lambda$ is the non-degenerate two form that is exact on the cotangent bundle $T^*M$, and the vector field $u^{(n)}$ is defined by
\begin{equation}
\text{i}_{u^{(n)}} \, \omega^{(n)} = -\text{d}H^{(n)},
\end{equation}
where
\begin{equation}
\omega^{(n)} \equiv \sum_{i=1}^{n}{\omega_i}.
\end{equation}
This equation simply states the the $N$-particle distribution form, $\rho^{(N)}$, is advected (that is deformed) by the vector field, $u^{(N)}$ on the manifold given by the cotangent bundle $T^*M^{(N)}$.  See Fig.~\ref{fig:advect}  for a pictorial representation of this general advection.  

Integrating this equation over the first $N$-$n$ particles gives the Generalized BBGKY Hierarchy,
\begin{equation}
\label{eqn:gbbgky}
\frac{\partial \rho^{(n)}}{\partial t} + \mathscr{L}_{u^{(n)}} \rho^{(n)} = - n_0 \int_{T^*M}{\mathscr{L}_{u^{(n)}_\text{int}} \rho^{(n+1)}},
\end{equation}
where the interaction velocity of the first $n$ particles with the $n$+$1$ particle is
\begin{equation}
u^{(n)}_\text{int} \equiv \sum_{i=1}^{n}{u_{i,n+1}}.
\end{equation}
This formalism is extremely general.  The distribution, $\rho$, can be interpreted as a general statistical distribution or the state of a quantum field theory.  For specific forms of $H$, the pull back of these equations can be shown to be:  Liouville equation, BBGKY hierarchy, Master equation, Vlasov equation, Boltzmann equation, multi fluid equations, Navier-Stokes equations, MHD equations, heat diffusion equation, radiation transport equations, quantum field equations, Schrodinger equation, Maxwell's equations, Newton's equations, etc.  This is why the Lipschitz continuity (that is continuity under diffeomorphism, deformation, or advection) is such a key property of the MST.

\begin{figure}[ht]
\center\includegraphics[width=20pc]{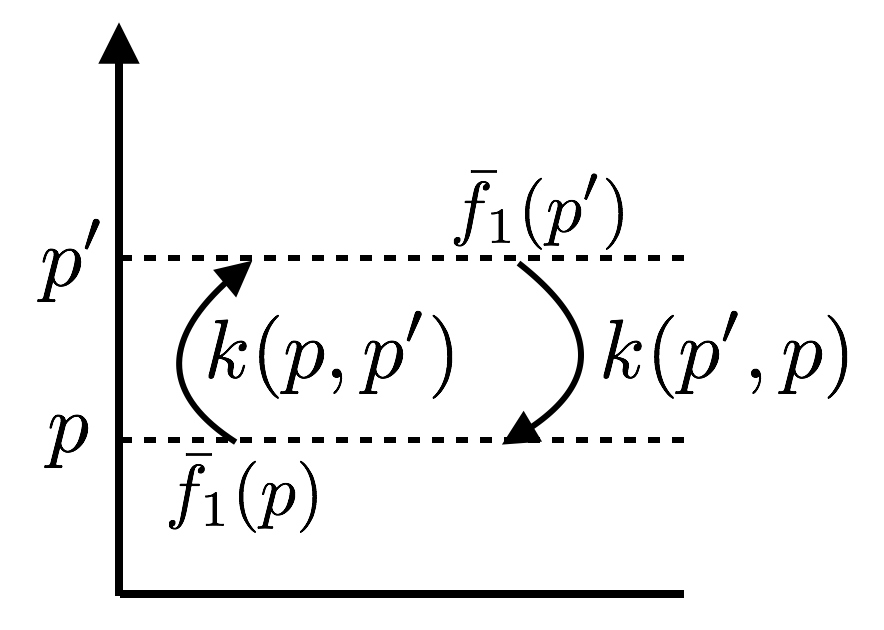}
\caption{\label{fig:tranrate} Transitions of the Generalized Master Equation, Eq.~\ref{eqn:gme}.}
\end{figure}

Using the ideas of Bogoliubov, we now pull back the Generalized BBGKY Hierarchy, Eq.~\ref{eqn:gbbgky}, and reduce it to a Generalized Master Equation.  We first pull back the equations for the $1$-particle distribution function, $f_1$, and the $2$-particle distribution function, $f_2$ and obtain
\begin{equation}
\label{eqn:bbgky1}
\frac{\partial f_1}{\partial t} + \{f_1,H_1\} = - n_0 \int{dp_2 \, dq_2 \, \{f_2,H_{12} \} }
\end{equation}
and
\begin{equation}
\label{eqn:bbgky2}
\frac{\partial f_2}{\partial t} + \{f_2,H_1+H_2+H_{12}\} = - n_0 \int{dp_3 \, dq_3 \, \{f_3,H_{13}+H_{23} \} }.
\end{equation}
The ideas of Bogoliubov are that $f_1$ relaxes at the dynamic rate, $\Omega$, to $\bar{f}_1$.  This $\bar{f}_1$ evolves at the collisional rate defined as
\begin{equation}
\frac{d \Omega / dt}{\Omega} \ll \Omega,
\end{equation}
which is much slower than the dynamic rate.  The $f_2$ relaxes at this collisional rate to $\bar{f}_2$.  This $\bar{f}_2$ then evolves at the correlation rate defined as
\begin{equation}
\frac{d^2 \Omega / dt^2}{\Omega^2} \ll \frac{d \Omega / dt}{\Omega} \ll \Omega,
\end{equation}
which is much slower than the collision rate.  We can now combine and reduce Eqns.~\ref{eqn:bbgky1} and ~\ref{eqn:bbgky2} using this separation of scales to the Generalized Master Equation,
\begin{equation}
\label{eqn:gme}
\frac{\partial \bar{f}_1(p)}{\partial t} - \frac{\partial \bar{f}_\text{source}(p)}{\partial t}  \backsim \int dp' \; \bar{f}_2(p',p)  - \bar{f}_2(p,p') =  \int dp' \; \bar{f}_1(p') \, k(p',p) - \bar{f}_1(p) \, k(p,p'),
\end{equation}
where the transition kernel is defined as
\begin{equation}
k(p,p') \equiv \frac{\bar{f}_2(p,p')}{\bar{f}_1(p)},
\end{equation}
and we have added a source term, $f_\text{source}$.  The right hand side of this equation has the simple interpretation shown in Fig.~\ref{fig:tranrate}.  The change in the probability is the integral of stuff coming into the state, $p$, from every state, $p'$, less the integral of the stuff going out of state, $p$, into every state, $p'$.  The state, $p$, can be the inverse scale, $1/\lambda$, the canonical momentum, the energy, the action, or the quantum numbers of the state.  Note that the average distribution function, $\bar{f}_1$ is considered constant over the dynamical scale in time and space.  This sets the scale for the local stationarity of the statistics, or the size of the neighborhood over which one can take the MST, or equivalently the size of the Father Wavelet.  This also highlights the need for the local stationarity of the MST;  this is in addition to the Lipschitz continuity already discussed.

Let's digress and discuss the importance of the Generalized Master Equation, as the fundamental equation of nonlinear dynamics and its relationship to emergent behavior and self organization.  First of all one should recognize $\bar{f}_1$ as the fundamental distribution function of the system.  The Generalized Master Equation, Eq.~\ref{eqn:gme}, gives the evolution this fundamental distribution.  Note that it can be written as an integrated difference between the $\bar{f}_2$'s.  Remember that $\bar{f}_2$ is the two-point correlation function.

The traditional way that the dynamics of Eq.~\ref{eqn:gme} is analyzed is via a linear analysis.  The $1$-particle distribution is linearized
\begin{equation}
\bar{f}_1(p,t) \approx f_0(p,t) + \delta f(p,t),
\end{equation}
where
\begin{equation}
\label{eqn:linasump}
\delta f / f_0 \ll 1,
\end{equation}
along with Eq.~\ref{eqn:gme} yielding a dispersion relation,
\begin{equation}
D(p,t) = 0.
\end{equation}
This dispersion relation can then be solved yielding the complex linear normal mode frequencies, $p_0 = k + i \gamma$.  This can be separated into the real oscillating frequency, $1/k$, and the imaginary stable or unstable growth rate, $\gamma$.

\begin{figure}[ht]
\center\includegraphics[width=25pc]{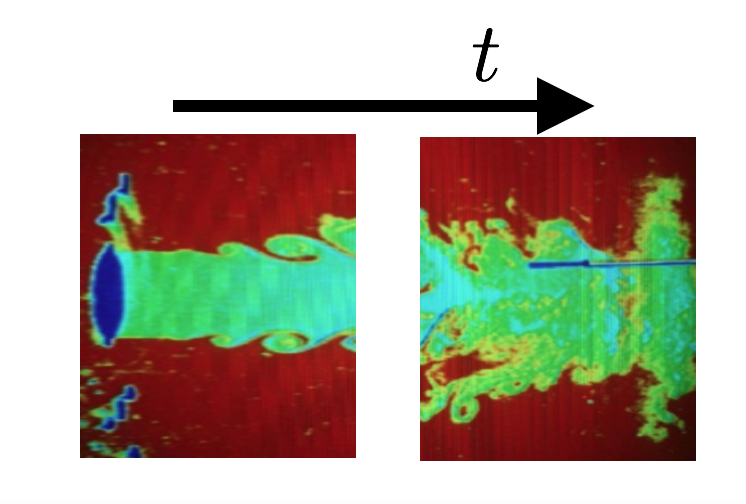}
\caption{\label{fig:khturb} Time evolution of Kelvin-Helmholtz shear instability of jet, as it undergoes a normal turbulent cascade.}
\end{figure}

\begin{figure}[ht]
\center\includegraphics[width=20pc]{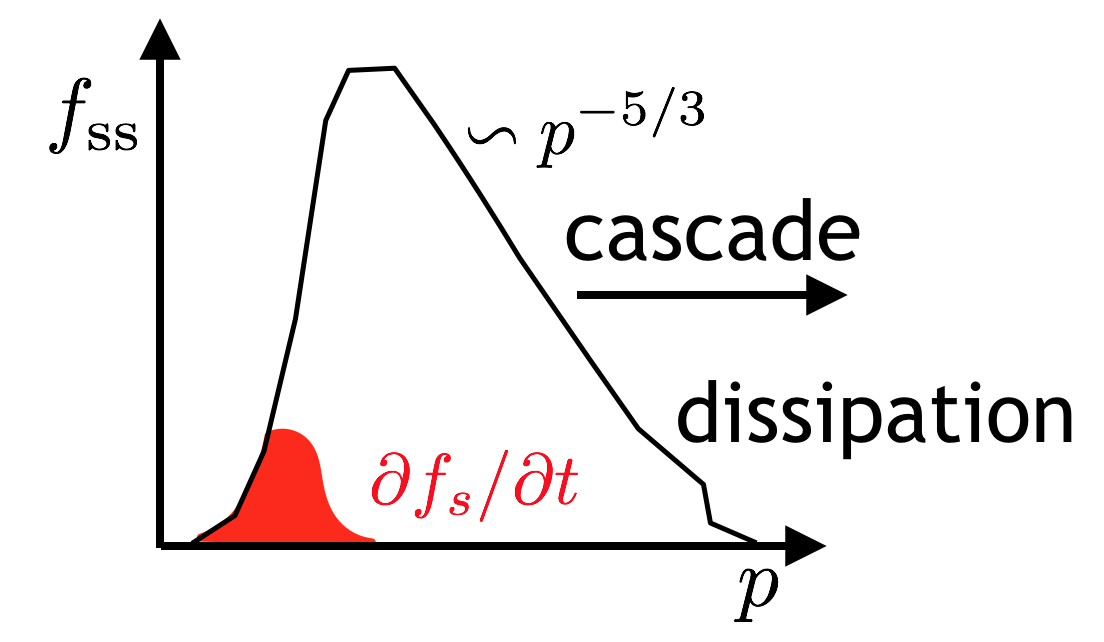}
\caption{\label{fig:cascade} Steady state spectrum of a normal cascade.}
\end{figure}

In contrast, as the system perturbation amplitude grows, it eventually enters the nonlinear regime where the linear assumption of Eq.~\ref{eqn:linasump} no longer applies.  The system now nonlinearly interacts, goes through a transient evolution and will approach a steady state,
\begin{equation}
f_\text{ss} (p) \equiv \lim_{t \to \infty}{\bar{f}_1(p,t)},
\end{equation}
which satisfies the equation
\begin{equation}
\int dp' \; f_\text{ss}(p) \, k(p,p') - f_\text{ss}(p') \, k(p',p) \backsim  \frac{\partial f_\text{source}(p)}{\partial t}.
\end{equation}
For most systems, such as 3D Navier-Stokes flow, the emergent behavior of the system or equivalently the steady state distribution under goes a normal cascade.  (See Fig.~\ref{fig:khturb}.)  That is to say that for a system being pumped around a specific frequency, the energy is transported from large to small scale in steady state and sets up the well known Kolmogorov scaling.  The energy is then dissipated at the small scale. (See Fig.~\ref{fig:cascade}.)  This is an emergent behavior with a very specific correlation structure.  The steady state, $f_\text{ss}(p)$, is encoded in the transition rates, $k(p,p') = \bar{f}_2(p,p') / \bar{f}_1(p)$.

\begin{figure}[ht]
\center\includegraphics[width=25pc]{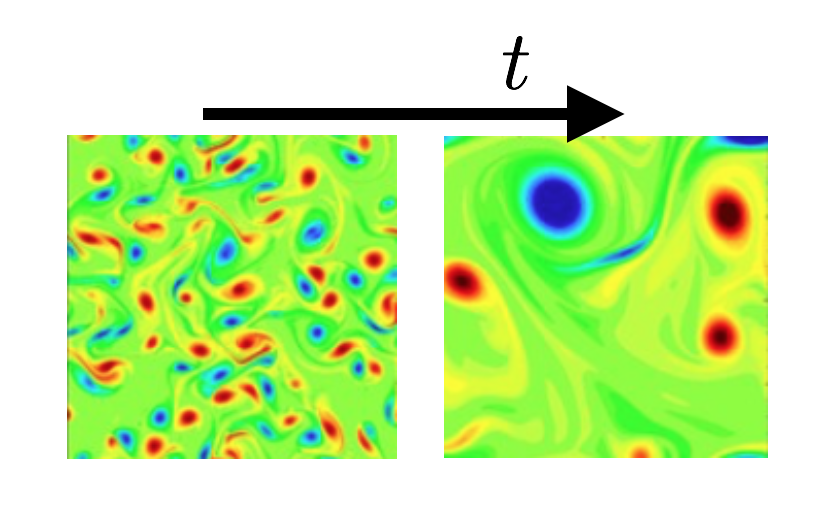}
\caption{\label{fig:nsturb}Time evolution of 2D Navier-Stokes turbulence as it undergoes an inverse cascade.}
\end{figure}

\begin{figure}[ht]
\center\includegraphics[width=20pc]{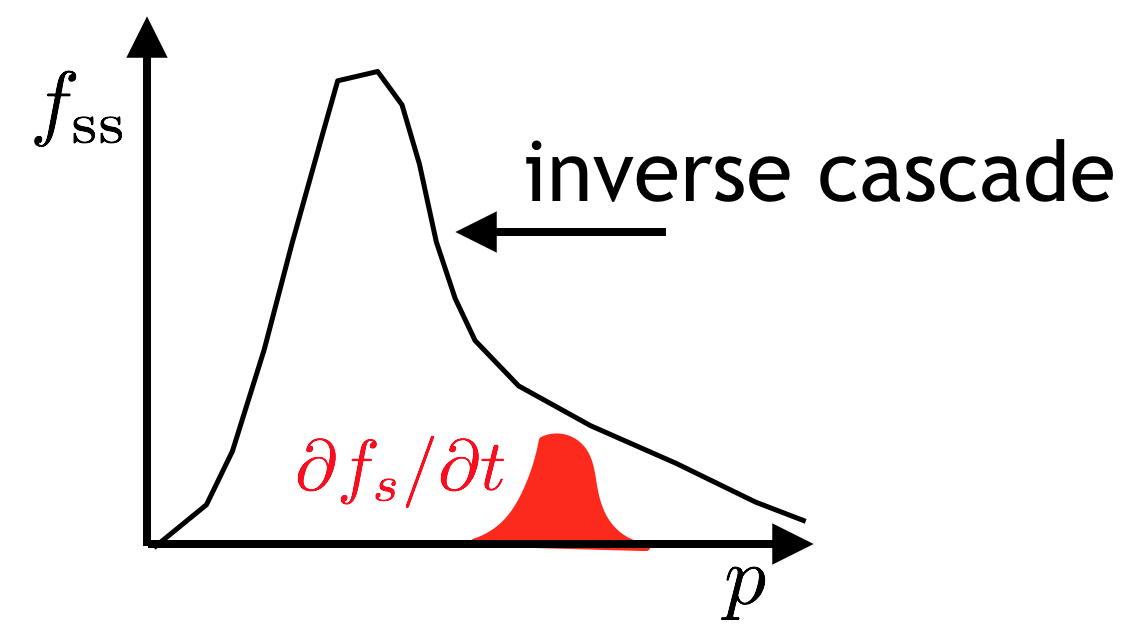}
\caption{\label{fig:invcascade} Steady state spectrum of an inverse cascade.}
\end{figure}

There is another class of systems that display a very different behavior.  The simplest example of such systems is 2D Navier-Stokes flow. (See Fig.~\ref{fig:nsturb}.)  Because the flow is constrained to 2D, the circulation can not un-twist.  It would need to go into 3D to do that.  As a result,
\begin{equation}
\text{total vorticity} = \int \nabla \times u \; d^2 x,
\end{equation}
is conserved.  The system then goes to a steady state where the vorticity is conserved, but the energy is minimized.  This also will be encoded in the transition rates, $k(p,p')$.  The system will approach a large scale, emergent behavior, that is called self organization.  The energy cascades from small scale to large.  This is called an inverse cascade.  (See Fig.~\ref{fig:invcascade}.)

\begin{figure}[ht]
\center\includegraphics[width=20pc]{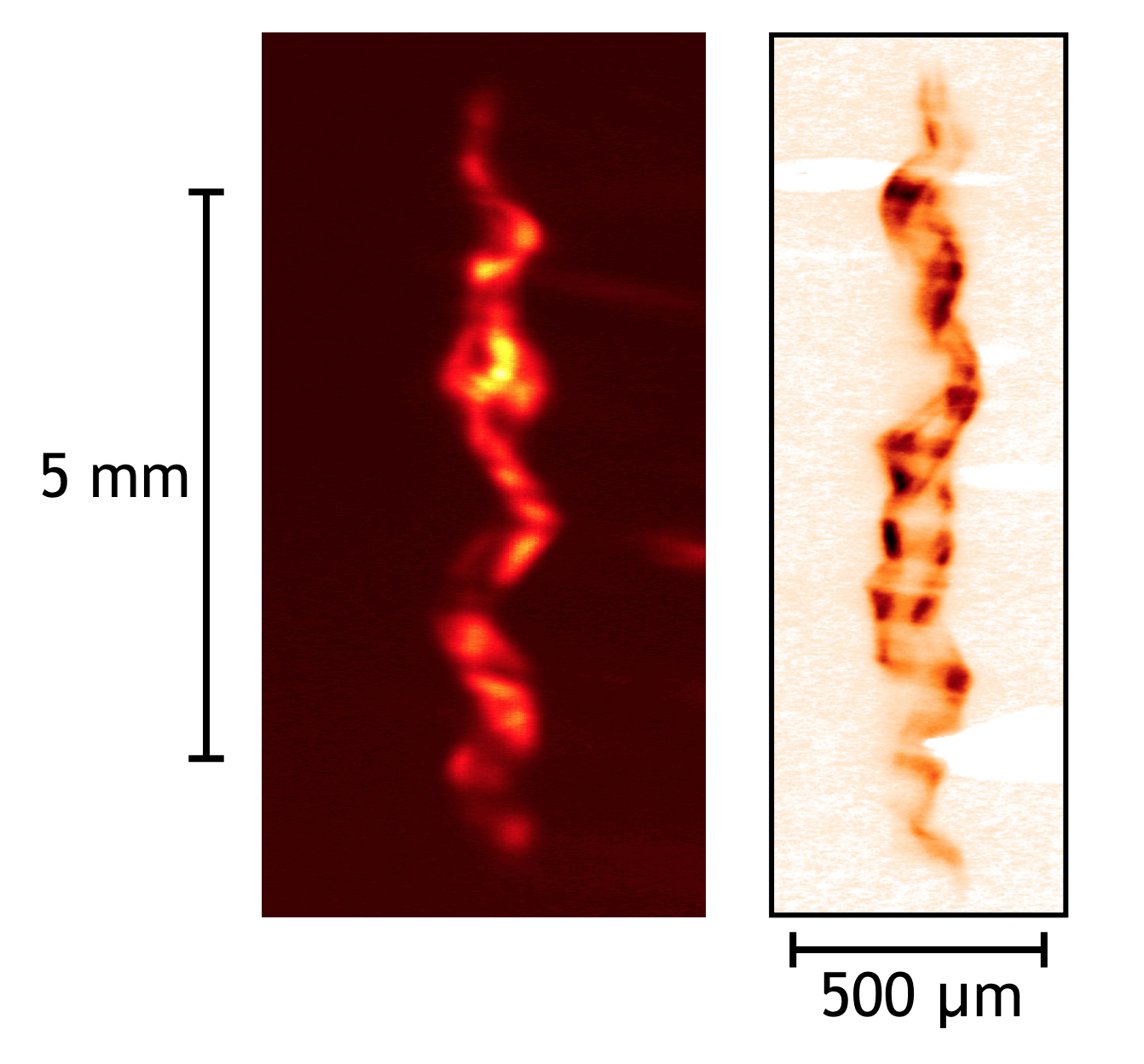}
\caption{\label{fig:maglifstag} Two stagnation, x-ray self emission, images of a MagLIF stagnation.}
\end{figure}

The most relevant case for MagLIF stagnation is 3D MHD evolution.  This case also undergoes and inverse cascade.  In this case, it is because of 
\begin{equation}
\text{total magnetic helicity} = \int A \cdot B \; d^3 x,
\end{equation}
and
\begin{equation}
\text{total cross helicity} = \int v \cdot B \; d^3 x,
\end{equation}
conservation.  Analogous to the 2D Navier-Stokes, the plasma evolves to a self organization (DNA-like double helix, see Fig.~\ref{fig:maglifstag}) as it implodes.

\begin{figure}[ht]
\center\includegraphics[width=30pc]{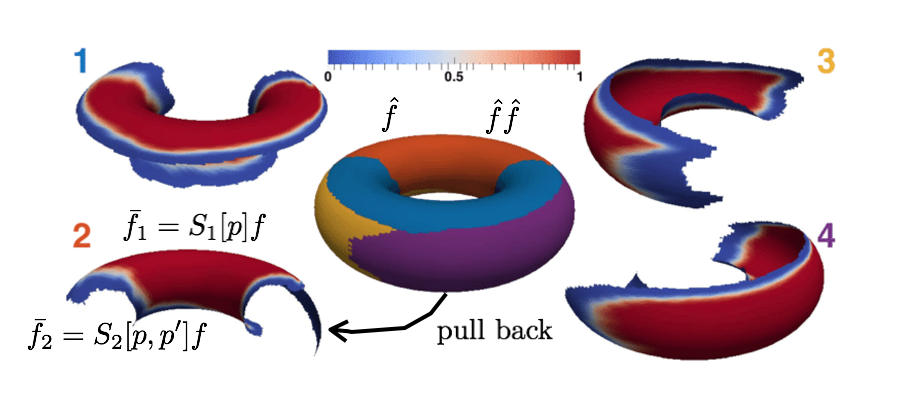}
\caption{\label{fig:punity} Partition of Unity on the manifold of a torus, $T^2$.}
\end{figure}

Now, we will make the connection between the quantities that appear in the Generalized Master Equation, $\bar{f}_1(p)$ and $\bar{f}_2(p,p')$, and the MST.  Remember that $\bar{f}_1$ and $\bar{f}_2$ encode the nonlinear steady state of the system, as well as the current state and its evolution.  We start by considering a piece of mathematics that dates to the early days of quantum mechanics in 1927 -- the Wigner-Weyl transformation \citep{wwt}.  This transformation takes operators to/from phase space, and faded into the recesses of history because of technical problems of mapping from the manifold structure on which the operators live to the $\mathbb{R}^n$ topology of phase space.  The origin of this technical problem is its solution -- a Modified Wigner-Weyl transformation needs to be constructed that is manifold safe.  Let us start by examining the structure of the forward Wigner map
\begin{equation}
\label{eqn:wignermap}
\tilde{W}[\hat{A}] \equiv \int{ds \; \psi^{*}_p (-s) \left< q+s \left| \hat{A} \right| q-s \right> \psi_p(s)} = A(q,p),
\end{equation}
where a Fourier kernel is used,
\begin{equation}
\psi_p(s) = e^{-i p \cdot s}.
\end{equation}
This is a bi-Fourier transform, since operators have both inputs and outputs, and both need to be transformed.  This transform also goes by the name of the bi-spectral transformation \citep{bst} in the modern machine learning literature.  The problem with this un-modified transformation is that the kernel has infinite support.  Integrations that are done on manifolds need to have a partition of unity, that is, patch functions.  Transformation kernels also need to have compact support.  We therefore propose the following modification to the Wigner-Weyl transformation.  Use the Mother Wavelet, $ \psi(x)$, as the kernel, and the Father Wavelet, $\phi(x)$, as the partition of unity.  (See Fig.~\ref{fig:punity}.)  We can now take the modified Wigner map of the density operator, which is called the Wigner function
\begin{equation}
\label{eqn:wignerfunc}
 \tilde{W}_f(q,p) \equiv \tilde{W}[\hat{\rho}] = \tilde{W}[ \, \left| f \right> \left< f \right| \,].
\end{equation}
Using the definition of the Wigner map given in Eq.~\ref{eqn:wignermap}, we find that the Wigner function equals
\begin{equation}
 \tilde{W}_f(q,p) = |f \star \psi_p |^2,
\end{equation}
using the terminology of the MST.  We can now compute $\bar{f}_1$ and $\bar{f}_2$ as
\begin{equation}
\bar{f}_1(p) \equiv  \text{E}(\tilde{W}[\hat{f}])= |f \star \psi_p | \star \phi = S_1[p]f
\end{equation}
and
\begin{equation}
\bar{f}_2(p,p') \equiv  \text{E}(\tilde{W}[\hat{f} \hat{f}])= ||f \star \psi_p | \star \psi_{p'}| \star \phi = S_2[p,p']f.
\end{equation}
The relationship is simple and profound.  The first order MST, $S_1[p]f$, is equal to, $\bar{f}_1(p)$, the state of the system.  The second order MST, $S_2[p,p']f$, is equal to $\bar{f}_2(p,p')$, how the state changes with time.  Together they specify the nonlinear dynamics of the system and the nonlinear steady state of the system.

\subsection{Field Theoretical:  S-matrix}
\begin{figure}[ht]
\center\includegraphics[width=20pc]{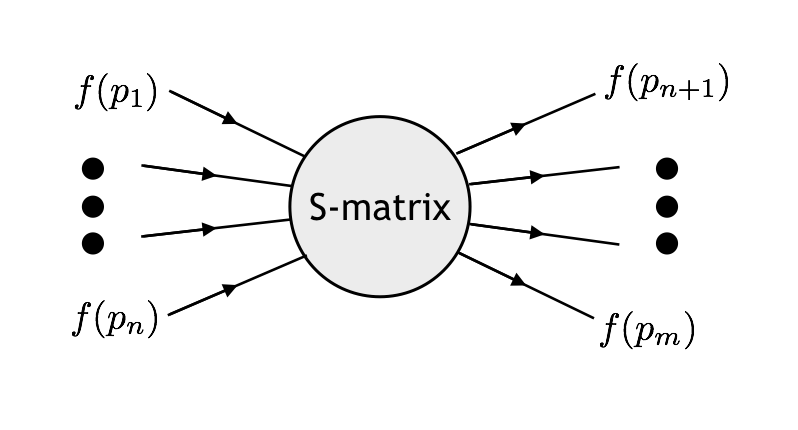}
\caption{\label{fig:smatrix} S-matrix which is the scattering cross section of $n$ particles into $m$-$n$ particles.}
\end{figure}

The connection of the MST to the mathematical constructs of field theory is simple and elegant.  Many of the details of this calculation that examines the MST as a new perspective on renormalization can be found in \citet{glinsky.11}.  As with the classical analysis of the previous section, we will find that the first order MST is related to the state of the system and the second order MST is related to the transition rate between states.  Let us start with the Lagrangian perspective and define the generating function, 
\begin{equation}
\label{eqn:zfunc}
Z[J]=N \int{[df(p)] \; \text{e}^{(i/\hbar) \, S_0[f(p)]+(i/\hbar) \int{dp \, J(p) \, f(p)}}},
\end{equation}
where $S_0[f(p)]$ is the action.  The connection to the canonical formalism is through the calculation of the Generalized Green's functions or equivalently the $m$-particle scattering cross sections (also known as the S-matrix, see Fig.~\ref{fig:smatrix}).  These are the functional Taylor coefficients of the generating function.  The S-matrix is
\begin{equation}
\label{eqn:smatrix}
S_m(\left|f\right>) \equiv \text{E}(T_p(\hat{f}(p_1) \dots \hat{f}(p_m)) \, F(f)) = ||f \star \psi_{p_1} | \dots \star \psi_{p_m}| \star \phi = \left. \frac{1}{Z[J]} \frac{\delta}{\delta J(p_1)}  \dots \frac{\delta}{\delta J(p_m)} Z[J] \right|_{J=0},
\end{equation}
where $f$ is a field, $F(f)$ is an ensemble of fields or the state of the system $\left|f\right>$, $T_p()$ is the $p$-ordered product operator, and $E()$ is the expected value operator.  Already from this expression the first order MST can be identified as the one particle S-matrix, and the second order MST can be identified as the two particle S-matrix.  As in the previous section on the classical interpretation, Eq.~\ref{eqn:smatrix} can be viewed as multiple field correlation functions.

\begin{figure}[ht]
\center\includegraphics[width=12pc]{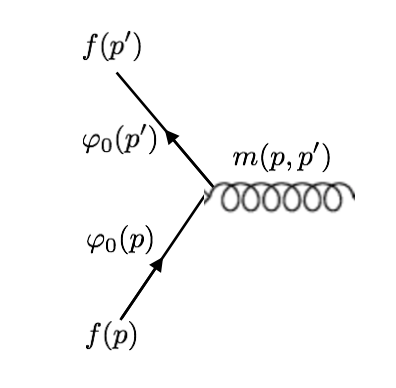}
\caption{\label{fig:fdiagram} Feynman diagram showing the physical significance of both $\varphi_0 (p) = S_1[p]f = \bar{f}_1(p)$ and $1/m(p,p') = S_2[p,p']f = \bar{f}_2(p,p')$.}
\end{figure}

This is still a bit abstract.  Let us define the effective action through a Legendre transformation,
\begin{equation}
S[\varphi(p)]= -\ln Z[J] + \int{dp \, J(p) \, \varphi(p)}.
\end{equation}
Expanding in $S$ and $\varphi$, it can be shown that
\begin{equation}
\label{eqn:smatrix1}
S_1(\left| f \right>) = |f \star \psi_p | \star \phi = \text{E}(\hat{f}(p) \, F(f)) = \left. \frac{1}{Z[J]} \frac{\delta Z[J]}{\delta J(p)} \right|_{J=0} = \varphi_0 (p)
\end{equation}
and
\begin{equation}
\label{eqn:smatrix2}
S_2(\left| f \right>) = ||f \star \psi_p | \star \psi_{p'} | \star \phi = \text{E}(\hat{f}(p) \,\hat{f}(p') \, F(f)) = \left. \frac{1}{Z[J]} \frac{\delta^2 Z[J]}{\delta J(p) \delta J(p')} \right|_{J=0} = \frac{1}{m(p,p')},
\end{equation}
where $\varphi_0 (p)$ is the classical action averaged over fluctuations as a function of the inverse renormalization scale, $p$, and $m(p,p')$ is the renormalization mass as a function of the initial and final inverse renormalization scales, $p$ and $p'$.  This is shown by the Feynman diagram (Fig.~\ref{fig:fdiagram}), where a field with action $\varphi_0 (p)$ is scattered by a particle of mass $m(p,p')$ into a field with action $\varphi_0 (p)$.

To summarize, the relationships of the classical and field theory interpretation of the scattering cross sections are:
\begin{equation}
\label{eqn:summary1}
S_1(p) = S_1(\left| f \right>) =  |f \star \psi_p | \star \phi = \bar{f}_1(p) = \varphi_0 (p)
\end{equation}
and
\begin{equation}
\label{eqn:summary2}
S_2(p,p') = S_2(\left| f \right>)  =  ||f \star \psi_p | \star \psi_{p'} | \star \phi  = \bar{f}_2(p,p') = 1 / m(p,p').
\end{equation}
For both Eq.~\ref{eqn:summary1} and Eq.~\ref{eqn:summary2}, the first term is a statement of the order of the MST, the second indicates the order of the S-matrix, the third is a practical description of how it is calculated, the fourth relates it to the dynamically and collisionally averaged statistical distribution functions, and the last relates it to the field theory quantities.  Remember that the transition rate is $k(p,p') = \bar{f}_2(p,p')/\bar{f}_1(p)$.

\section{Conclusions}\label{sec:conclusion}
We have demonstrated a full machine learning pipeline which we have applied to characterize the morphology of experimental and simulated MagLIF plasma stagnation images. We demonstrated that the MST provides a convenient basis in which to project out discrepancies between simulated and experimental images. Furthermore, our method demonstrates that, when projected onto our double helix model, we are in some cases able to differentiate the morphology of different experiments. Particularly, by providing a characterization of the uncertainty in our predictions, we are able to conduct statistical hypothesis testing around discrepancies in morphology in experiment-to-experiment and experiment-to-simulation settings.  At the very least we are able to interpret the images, as well as difference between images, in terms of the parameters of the double helix model and the uncertainties in the estimates of those parameters.

We have also developed theory that connects the transformation to the causal dynamics of physical systems.  This has been done from the classical kinetic perspective, where the MST are expected values of the generalized Wigner-Weyl transformations of the density operator.  And, has been done from the field theory perspective, where the MST is the generalized Green's function, or S-matrix of the field theory in the scale basis.  From the classical perspective, the first order MST is the one particle distribution function, averaged over the fast dynamical time scale, and the second order MST is simply related to the fully nonlinear transition rate from one scale to another.  The first gives the current state, and the second gives the nonlinear evolution of the system.  From the field theory perspective, the first order MST is the classical action averaged over fluctuations as a function of the inverse scale, and the second order MST is the scattering cross section from an initial to a final scale.  The first again gives the current state of the system, and the second gives the inverse mass of the field boson that mediates the field interaction and scatters the field, thereby evolving the field.  What is required of the system is that it is causal.  Equivalently, a Lagrangian for the system can be written down, and therefore the system will be evolved according to an action principle.  This leads to a generalized advection by a Lie derivative, in the classical case, and by dynamical paths weighted by the exponential of their actions, in the field theory case.
    
Therefore, it is no surprise that the MST is a good metric for nonlinear systems.  It encodes both the initial state and the dynamics (transition rates) between states.  This explains why the first order MST, or the Fourier transformation, are not sufficient to uniquely identify the systems.  They only encode the the initial state.  There are other technical details that cause problems with the Fourier Transform, unless there are rotational symmetries in the physical system.  It is also the reason a system that encodes finite information is fully identified by the first and second order MST.  This comes about because of a statistical realizability theorem  -- either the distribution stops at second order or it must continue to all orders.  Since the information is finite, the distribution must stop at second order.  One can reason to this since knowing the S-matrix, that is the MST, is equivalent to knowing the Lagrangian of the field theory.
    
What is even more important about the MST is the connection to the dynamical evolution of the physics.  If one includes the evolution coordinate, that is time, in the transformation, the second order MST directly, and with no further transformation, gives the transition kernel of the dynamics.  This is independent of the current state, that is the first order MST.  Given an ensemble of example states that sufficiently sample the transition kernel, one has fully characterized the physical system and should be able to evolve any state forward in time, as given by the initial first order MST.  That is the MST is the perfect coordinate system in which to learn, identify, and propagate the dynamics.
    
The MST has been implemented in an efficient (GPU accelerated), yet flexible, Python framework based on Keras/Tensorflow.  This package includes 1D, 2D and 3D transformations along with visualization.  It supports, through adjoints, the inversion of the transformation.  The software is open source and distributed through PyPI as the BluSky project.

\section{Future Research Directions}\label{sec:future}
The research of this LDRD has developed both the mathematical formalism and efficient Python classes that will be the foundation of significant future application.  There are several ways that the MST will be used.  First, it will be used in the Bayesian Data Assimilation Engine that is under development for the analysis of MagLIF and other related experiments on the Z-Machine at Sandia.  Specifically, much of the data that is generated by Z is in the form of images and signals that have stochastic character that is not being quantitatively included in the objective functions.  The MST is a metric of this stochastic character that will enable this to be a further constraint on the analysis, thereby improving what can be estimated from the experiments.
 
Second, because of the relationships established by this research between the MST and the evolution of causal physical systems, the MST could be used as a coordinate system to numerically integrate physical dynamics.  This could be using traditional techniques or by using modern data science techniques, such as machine learning how to numerically integrate physical dynamics in the MST representation.  Given that there is emergent nonlinear behavior presenting itself during implosions on Z, this connection also indicates that the MST representation would be a good space in which to theoretically analyze the emergent behavior and identify phase transitions, that is bifurcations, of the emergent behavior.  Finally, there is great promise for using the MST as the coordinate system when fast surrogate models are being built with modern data science techniques for forward models, such as our rad-MHD simulations, that currently take days to weeks to generate one data point.

There are several improvements that can be made to the analysis done with the MST over the course of the last three years.  We note that our approach has made a number of reasonable assumptions. For example, we assume a model which exhibits a significant amount of symmetry that may not be realized in experiment. This may be lifted to an extent by modifying the analytical model. However, even when experimental images show strong asymmetry, the predicted morphology parameters demonstrate reasonable behavior. We have also assumed a linear relationship between the MST coefficients and helical model parameters. This assumption may be lifted by using a nonlinear regression method such as a neural network.  We have also not incorporated true 3D helical structure and correlation in our analysis.  We have worked with a 2D parameterization of the helical structure, not a 2D projection of a 3D structure.  Lifting some of these assumptions will likely reduce the uncertainty in our parameter inferences, making the method even more useful in characterizing morphology differences.  This could lead to physical insights regarding mechanisms creating particular plasma configurations and to improvements in target design. Finally, there are still many experimental images which are too noisy, or exhibit other artifacts which preclude our ability to get reliable morphology estimates. We are currently exploring additional machine learning methods, such as data augmentation, which may reduce the impact of experimental noise on our morphology regression. We will continue to pursue these ideas and present them in future work.

\begin{acknowledgments}
We would like to thank St\'ephane Mallat for many useful discussions, suggestions, and providing a computer software implementation of his Scattering Transformation.  This research was funded by the Sandia National Laboratories' Laboratory Directed Research and Development (LDRD) program.  Sandia National Laboratories is a multimission laboratory managed and operated by National Technology and Engineering Solutions of Sandia LLC (NTESS), a wholly owned subsidiary of Honeywell International Inc., for the U.S. Department of Energy's National Nuclear Security Administration (NNSA) under contract DE-NA0003525.  This paper describes objective technical results and analysis. Any subjective views or opinions that might be expressed in the paper do not necessarily represent the views of the U.S. Department of Energy or the United States Government.  The data that support the findings of this study are available from the corresponding author upon reasonable request.  This is a Sandia National Laboratories Technical Report, SAND2019-11910.
\end{acknowledgments}

\bibliography{Moore_PoP2019.bib}

\end{document}